\documentclass[useAMS,usenatbib]{mn2e}

\usepackage[normalem]{ulem} % to cross out part of the text

\usepackage{graphicx}
\usepackage{hyperref}
\usepackage{color,epsfig}
\usepackage[utf8]{inputenc}
\usepackage{verbatim}
\usepackage{lmodern}
\usepackage{amssymb}

\newcommand{\be}{\begin{equation}}
\newcommand{\ee}{\end{equation}}

\newcommand*{\eqref}[1]{(\ref{#1})}

\def\msun{{\,{\rm M}_\odot}}
\def\rsun{{\,{\rm R}_\odot}}
\def\gcm3{\, \rm g \, cm^{-3}}

\def\rt{R_{\rm t}}
\def\rg{R_{\rm g}}
\def\rp{R_{\rm p}}
\def\ra{R_{\rm a}}
\def\mh{M_{\rm h}}
\def\va{v_{\rm a}}
\def\mstar{M_{\star}}
\def\rstar{R_{\star}}
\def\pstar{P_{\star}}
\def\astar{a_{\star}}

\def\rcirc{R_{\rm circ}}
\def\lcirc{L_{\rm circ}}
\def\tcirc{t_{\rm circ}}
\def\ncirc{n_{\rm circ}}
\def\tvisc{t_{\rm visc}}
\def\tcool{t_{\rm cool}}
\def\tdiff{t_{\rm diff}}
\def\tmin{t_{\rm min}}
\def\ecrit{e_{\rm crit}}

\def\mp{m_{\rm p}}
\def\sigmat{\sigma_{\rm T}}
\def\mdot{\dot{M}}
\def\mdotedd{\dot{M}_{\rm Edd}}
\def\ledd{L_{\rm Edd}}
\def\erg{\rm erg}

\title[Disc formation from TDEs]{Disc formation from tidal disruptions of stars on eccentric orbits by Schwarzschild black holes}
\author[Clément Bonnerot, Elena M. Rossi, Giuseppe Lodato and Daniel J. Price]{Clément Bonnerot$^{1}$\thanks{E-mail: bonnerot@strw.leidenuniv.nl}, Elena M. Rossi$^{1}$, Giuseppe Lodato$^{2}$ and Daniel J. Price$^{3}$\\
$^{1}$Leiden Observatory, Leiden University, PO Box 9513, 2300 RA, Leiden, the Netherlands\\
$^{2}$Dipartimento
 di Fisica, Università Degli Studi di Milano, Via Celoria, 16, Milano, 20133, Italy\\
$^{3}$Monash Centre for Astrophysics and School of Mathematical Sciences, Monash University, Clayton, Vic 3800, Australia
}
\begin{document}

\date{ Accepted ?. Received ?; in original form ?}

\pagerange{\pageref{firstpage}--\pageref{lastpage}} \pubyear{2014}

\maketitle

\label{firstpage}

\begin{abstract}
The potential of tidal disruption of stars to probe otherwise quiescent supermassive black holes cannot be exploited, if their dynamics is not fully understood. So far, the observational appearance of these events has been derived from analytical extrapolations of the debris dynamical properties just after disruption. By means of hydrodynamical simulations, we investigate the subsequent fallback of the stream of debris towards the black hole for stars already bound to the black hole on eccentric orbits. We demonstrate that the debris circularize due to relativistic apsidal precession which causes the stream to self-cross. The circularization timescale varies between 1 and 10 times the period of the star, being shorter for more eccentric and/or deeper encounters. This self-crossing leads to the formation of shocks that increase the thermal energy of the debris. If this thermal energy is efficiently radiated away, the debris settle in a narrow ring at the circularization radius with shock-induced luminosities of $\sim 10-10^3 \, \ledd$. If instead cooling is impeded, the debris form an extended torus located between the circularization radius and the semi-major axis of the star with heating rates $\sim 1-10^2 \, \ledd$. Extrapolating our results to parabolic orbits, we infer that circularization would occur via the same mechanism in $\sim 1$ period of the most bound debris for deeply penetrating encounters to $\sim 10$ for grazing ones. We also anticipate the same effect of the cooling efficiency on the structure of the disc with associated luminosities of $\sim 1-10 \, \ledd$ and heating rates of $\sim 0.1-1 \, \ledd$. In the latter case of inefficient cooling, we deduce a viscous timescale generally shorter than the circularization timescale. This suggests an accretion rate through the disc tracing the fallback rate, if viscosity starts acting promptly.
\end{abstract}


\section{Introduction}

Most galaxies have been found to contain a supermassive black hole (SMBH) at their centre orbited by stars. If one of these stars wanders within the tidal radius of the black hole, the tidal force of the black hole exceeds the star's self-gravity and the star is torn apart. Such an event is called a tidal disruption event (TDE). After the disruption, the stellar debris evolve to form an elongated stream of gas that falls back towards the black hole. In the standard picture of TDEs, these debris then circularize to form an accretion disc where they accrete viscously emitting a thermal flare mainly in the UV to soft X-ray band \citep{lodato2011}. This flare could also be accompanied by a radio signal associated to a relativistic jet originating from the inner region of the disc \citep{metzger2011}. A handful of candidate TDEs have been detected so far in these bands \citep[e.g.][]{esquej2008,komossa2004,gezari2009,van_velzen2011,zauderer2011}.

TDEs are powerful tools to detect SMBHs in otherwise quiescent galaxies. Furthermore, they can in principle be used to estimate the properties of both the black hole and the disrupted star and to probe the physics of accretion and relativistic jets. In practice, deriving such constraints from observations is challenging as it requires a precise understanding of the dynamics of TDEs. The latter has been the focus of many analytical and numerical investigations undertaken since the 80s. A distinctive feature of the pioneering works by \citet{lacy1982}, \citet{rees1988}, \citet{evans1989} and \citet{phinney1989} is a $t^{-5/3}$ decrease of the rate at which the stellar debris fall back towards the black hole.\footnote{This rate was later shown to be dependent on the stellar structure \citep{lodato2009,guillochon2013}.} The same slope is generally fitted to the observed TDE light curves \citep{esquej2008,gezari2008,gezari2009,cappelluti2009} although it is only expected in the X-ray band \citep{lodato2011}. It assumes that the accretion rate, and therefore the luminosity, traces the fallback rate of the debris. In turn, this requires that the debris circularize to form a disc and that this disc is accreted faster than it is fed from the fallback stream. However, the mechanism driving the circularization is still unknown.

Various effects are likely to be involved in this process, whose common feature is to dissipate the kinetic energy of the debris, injecting a large amount of thermal energy into the newly formed disc. However, the efficiency of this energy transfer is certainly dependent on the mechanism considered. The main candidates are the following \citep{evans1989,kochanek1994}:

\begin{enumerate}
\item \textit{Pancake shock}: As the star is disrupted, part of its material is accelerated out of the initial orbital plane. As a result, the debris inside the stream have a range of inclinations. Their orbits are therefore vertically focussed and intersect the orbital plane close to pericentre. At this point, the stream is strongly compressed, leading to the formation of a pancake shock.
\item \textit{Self-crossing}: When they reach pericentre, the debris experience changes of their apsidal angles driven by relativistic apsidal precession or hydrodynamical effects. This can lead to the self-crossing of the stream: as the leading parts move away from the black hole after pericentre passage, they collide with the part that is still falling back, generating shocks.
\item \textit{Shearing}: As the stream comes back to pericentre, the orbits of the debris are radially focussed due to their large range of apocentres but small range of pericentres. This effect is similar to a passage into a nozzle. The debris experience shearing at this point as they have a range of apsidal angles and eccentricities induced by the disruption of the star. This effect is enhanced by relativistic apsidal precession: as they have a range of pericentre distances, the orbits of the debris precess by different angles, leading to further shearing.

\end{enumerate}

If an accretion disc forms from the debris, its structure and evolution are two additional uncertainties in the models of TDEs. They depend on the relative efficiency of three processes \citep{evans1989}: circularization, viscous accretion and radiative cooling. Denoting the timescales of these processes by $\tcirc$, $\tvisc$ and $\tcool$ respectively, three limiting regimes are to be expected. In the case $\tvisc < \tcirc$, viscosity may be important during the circularization process. If instead $\tvisc > \tcirc$, accretion begins only once a disc is formed. In this case, if $\tcool < \tcirc$, the disc cools during its formation and is therefore geometrically thin. If instead $\tcirc < \tcool$, the disc puffs up while it forms due to its excess of thermal energy. Many authors have assumed that the disc is geometrically thin or slim \citep{strubbe2009,cannizzo2009,cannizzo2011,lodato2011,shen2014} using the standard $\alpha$ parametrization \citep{shakura1973}. Other investigations considered the possibility of a geometrically thick disc \citep{loeb1997,coughlin2014}.

Numerical simulations of tidal disruptions \citep{evans1989,lodato2009} have often used smoothed particle hydrodynamics (SPH) primarily because of its ability to deal with large regions of space devoid of gas. This technique is also well suited to simulate the subsequent fallback of the debris towards the black hole. The computational cost of such a simulation scales with the total number of SPH particles used to model the stream. To accurately follow their evolution, each part of the stream must contain enough particles. This condition is hard to fulfill for a long stream because the SPH particles are spread on a large volume. The length of the stream increases with the stellar to black hole mass ratio $q$ and with the eccentricity $e$ of the star. The typical values of these parameters are $q=10^6$ and $e=1$. This extreme value of $e$ comes from the fact that most disrupted stars are scattered from the sphere of influence of the SMBH, which is much larger than the tidal radius \citep{frank1976}. For the Milky Way, the ratio of these two distances is $\sim10^5$. However, as noted by \citet{ayal2000}, following the fallback of the debris for these typical values of $q$ and $e$ numerically is a computational burden. In their simulation, the leading part of the stream is composed of very few SPH particles that come back almost one by one towards the black hole. As a result, the evolution of these particles cannot be followed accurately. Even if they use a relatively low number of particles ($N=4295$), this issue is so extreme that it is likely to persist for larger particle numbers. The high computational cost of such a simulation is not limited to SPH but generalizes to other computational techniques.

In the few other investigations of this problem, either $q$ or $e$ has therefore been lowered. The first option has been chosen by several authors \citep{ramirez-ruiz2009,guillochon2014} who investigated $q=10^3$ and $e=1$. The physical motivation in this case is the tidal disruption of a star by an intermediate mass black hole. Both pancake shock and shearing were shown to be inefficient at circularizing the debris in this case. Instead, the main effect seen in these simulations is the expansion of the stream caused by its pericentre passage. Circularization is more likely driven by self-crossing either induced by this expansion or by relativistic apsidal precession. \citet{hayasaki2013} made the second choice, as in the present paper, which corresponds to tidal disruptions of bound stars. They considered $q=10^6$ and $e=0.8$ and found that self-crossing driven by relativistic precession efficiently drives the circularization of the debris in this case.

Several mechanisms are able to put a star on a bound orbit entering the tidal radius of a SMBH. The first involves a unequal mass SMBH binary interacting with a surrounding stellar cusp. In this situation, bound stars can be scattered inside the tidal radius of the primary black hole through a combination of Kozai interactions and close encounters with the secondary black hole \citep{chen2009,chen2011}. The second mechanism takes place during the coalescence of a SMBH binary. Due to the anisotropic emission of gravitational waves in this phase, the remnant black hole undergoes a kick. It can lead to close encounters between the black hole and the surrounding stars leading to their tidal disruptions \citep{stone2011}. Furthermore, mean-motion resonances can pull surrounding stars inwards during the final phase of the SMBH merger \citep{seto2010} enhancing the rate of tidal disruptions of bound stars by this mechanism. The third mechanism takes place after the tidal separation of a stellar binary by a black hole, which places one of the components on an eccentric orbit. Due to both encounters with the other stars orbiting the SMBH and gravitational wave emission, its orbit shrinks and circularize. Under certain conditions, it can then be tidally disrupted by the SMBH on a bound orbit \citep{amaro-seoane2012}.

In this paper, we use SPH simulations to investigate the circularization process leading to the formation of the disc. We also characterize the structure and evolution of this disc. Relativistic apsidal precession, involved in the circularization process, is treated exactly while most previous studies \citep{ramirez-ruiz2009,guillochon2014,hayasaki2013} used an approximate treatment of this effect. The parameters that are likely to play a significant role in the circularizaton process have also been varied. Two extreme cooling efficiencies, encoded in the equation of state (EOS) of the gas, have been considered. The effect of the orbit of the star has also been examined, by varying both its penetration factor and its eccentricity. Our results demonstrate that circularization is a fast process, happening in a few orbits of the debris around the black hole, and driven mostly by relativistic apsidal precession. They also confirm the expected effect of the cooling efficiency on the structure of the resulting disc. In addition, we found different channels of circularization for different orbits of the star. In particular, by increasing its eccentricity, we got insight into the circularization process at work in the standard parabolic case.

This paper is organized as follows. In Section \ref{simulations}, we describe the SPH simulations performed. The results of these simulations are presented in Section \ref{results} varying the following parameters: the gravitational potential, the EOS, the depth of the encounter and the eccentricity of the star. The effect of the resolution on these results is also examined. These results are discussed in Section \ref{discussion} and compared to other studies by \citet{shiokawa2015} and \citet{hayasaki2015} that appeared at the same time as this paper.

\begin{figure}
\epsfig{width=0.47\textwidth, file=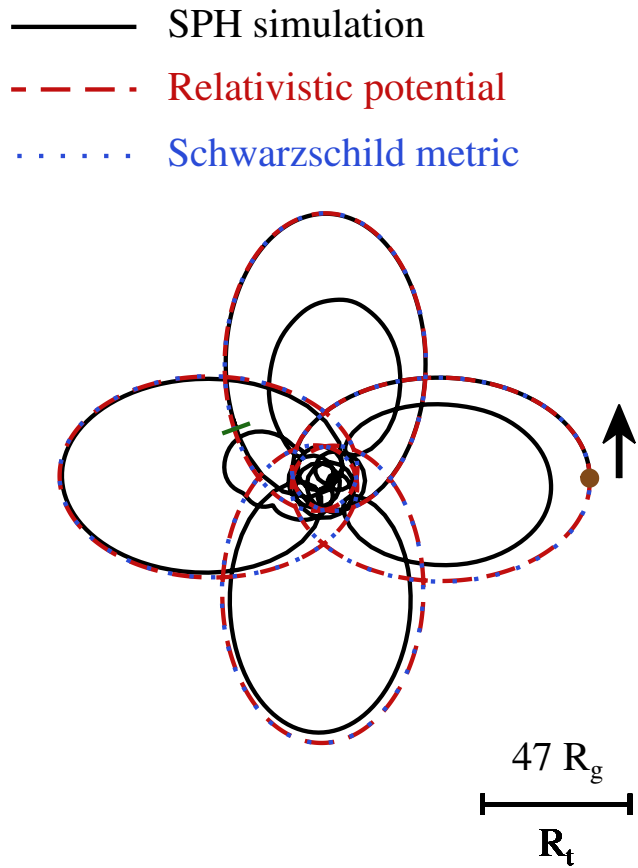}
\caption{Trajectory of the centre of mass of the gas in the simulation corresponding to model RI5e.8 (black solid line) compared to that of a test particle on the same orbit in the relativistic potential (red dashed line) and in the Schwarzschild metric (blue dotted line). All the trajectories start at the initial position of the star indicated by the brown point on the right of the figure. The black arrow specifies the direction of motion. The trajectory of the test particle in the relativistic potential and in the Schwarzschild metric is followed for 4 orbits. That of the centre of mass of the gas in the simulation is followed during the disruption of the star and the fallback of the debris until $t/\pstar = 8$. The transition between the disruption and the fallback phase, treated with two different codes, is indicated by the green dash perpendicular to this trajectory. After a few pericentre passages, the motion of the gas is affected by hydrodynamical effects and the trajectory of its centre of mass differs from that of a test particle.}
\label{fig1}
\end{figure}

\section{SPH simulations}
\label{simulations}

We investigate the tidal disruption of a star of mass $\mstar = 1 \msun$ and radius $\rstar = 1 \rsun$ by a non-rotating black hole of mass $\mh = 10^6 \msun$. In this configuration, the tidal radius is $\rt = \rstar (\mh/\mstar)^{1/3}=100 \rsun$. Several initial elliptical orbits of the star are considered. They have different pericentre distances $\rp$ defined via the penetration factor $\beta=\rt/\rp$ by setting $\beta = 1$ and $\beta = 5$. It corresponds to pericentre distances $\rp = 100\rsun$ and $\rp =20 \rsun$ respectively. For $\beta = 1$, only an eccentricity $e=0.8$ is considered. For $\beta = 5$, two different values $e = 0.8$ and $e =0.95$ are investigated. For these orbits, all the debris produced by the disruption stay bound to the black hole. This is generally the case if $e < \ecrit = 1-(2/\beta)(\mh/\mstar)^{-1/3}$, where $\ecrit = 0.996$ and $0.98$ for $\beta=5$ and $\beta=1$ respectively. The semi-major axis $a_{\star}=\rp/(1-e)$ of these orbits ranges between $100 \rsun$ and $500 \rsun$, corresponding to orbital periods $\pstar = 2 \pi \left({G \mh} / a^3_{\star} \right)^{-1/2}$ between $2.8 \, \rm h$ and $31 \, \rm h$.

Both the disruption of the star and the subsequent fallback of the debris towards the disruption site are simulated using SPH. To increase efficiency, the simulations of these two phases are performed separately and make use of two different codes, each adapted to a specific phase. For the disruption phase, we use a code that takes into account self-gravity \citep{bate1995}. For the fallback phase, we do not consider self-gravity and make use of the highly efficient code PHANTOM \citep{price2010,lodato2010} that is optimized for studying non-self-gravitating problems\footnote{Self-gravity is now implemented into PHANTOM. However, it was not available at the beginning of this work.}. This choice is legitimate because the self-gravity force is only needed during the disruption phase where it opposes the tidal force of the black hole. In the fallback phase, the gravitational interactions between the debris is negligible compared to their hydrodynamical interactions and the tidal force of the black hole. The disruption phase is followed until the most bound debris come back to pericentre. Their properties are recorded at this point and constitute the initial conditions for the simulation of the fallback phase. As this paper's primary focus is on the fallback of the debris, the disruption phase is only simulated to get these initial conditions and not discussed in detail.

The code used for the disruption phase has already been adopted by \citet{lodato2009} to simulate tidal disruptions of stars on parabolic orbits. In the present paper, we follow the same procedure and numerical setup but consider elliptical orbits instead. Also using the same method, we model the star as a polytropic sphere with $\gamma=5/3$ containing 100K particles. However, as explained in subsection \ref{convergence}, the fallback phase is simulated at a higher resolution with the stream of debris containing 500K particles.

In order to resolve the shocks, the code used for the fallback phase, PHANTOM, includes the standard artificial viscosity prescription that depends on the parameters $\alpha^{AV}$ and $\beta^{AV}$. In addition, the \citet{morris1997} switch is implemented to reduce artificial dissipation away from shocks. To this end, $\alpha^{AV}$ is allowed to vary between two values $\alpha^{AV}_{\rm min}$ and $\alpha^{AV}_{\rm max}$ according to a source and decay equation. In this paper, we use $\alpha^{AV}_{\rm min} =0.01$, $\alpha^{AV}_{\rm max}=1$ and $\beta^{AV}=2$.

The orbits considered for the star have pericentre distances comparable with the gravitational radius $\rg = G \mh / c^2$ of the black hole. More precisely, $\rp = 47 \rg$ and $\rp = 9.4 \rg$ for $\beta=1$ and $\beta=5$. Therefore, relativistic effects must significantly affect the motion of the gas when it passes at pericentre. One of these effects is relativistic apsidal precession, a mechanism involved in the circularization process causing the orbit of a test particle to precess at each pericentre passage by a given precession angle. For the orbits considered, the precession angle varies between 13.5 and 89.7 degrees for $\beta=1$ and $\beta=5$ respectively. In order to investigate the influence of relativistic effects, the black hole is modelled by an external potential that is either Keplerian or relativistic both in the disruption and the fallback phase. In the following, the variables are labelled by the letter K if they are computed in the Keplerian potential and by the letter R if they are computed in the relativistic potential. These potentials are respectively given by
\be
\Phi^{\rm K}=-\frac{G\mh}{R},
\ee
\be
\Phi^{\rm R}=-\frac{G\mh}{R}-\left(\frac{2 \rg}{R - 2 \rg}\right) \left[\left(\frac{R - \rg}{R - 2 \rg}\right) v^2_{\rm r} + \frac{v^2_{\rm t}}{2} \right],
\label{pot_rel}
\ee
where $v_{\rm r}$ and $v_{\rm t}$ are the radial and tangential velocity of a test particle respectively. We adopt this relativistic potential, designed by \citet{tejeda2013}, because it contains an exact treatment of apsidal precession around a non-rotating black hole. In other words, the value of the precession angle for a test particle on a given orbit is the same in this potential and in the Schwarzschild metric. This is an improvement compared to most previous investigations that used an approximate treatment of this effect \citep{ramirez-ruiz2009,guillochon2014,hayasaki2013}.

\begin{figure*}
\epsfig{width=\textwidth, file=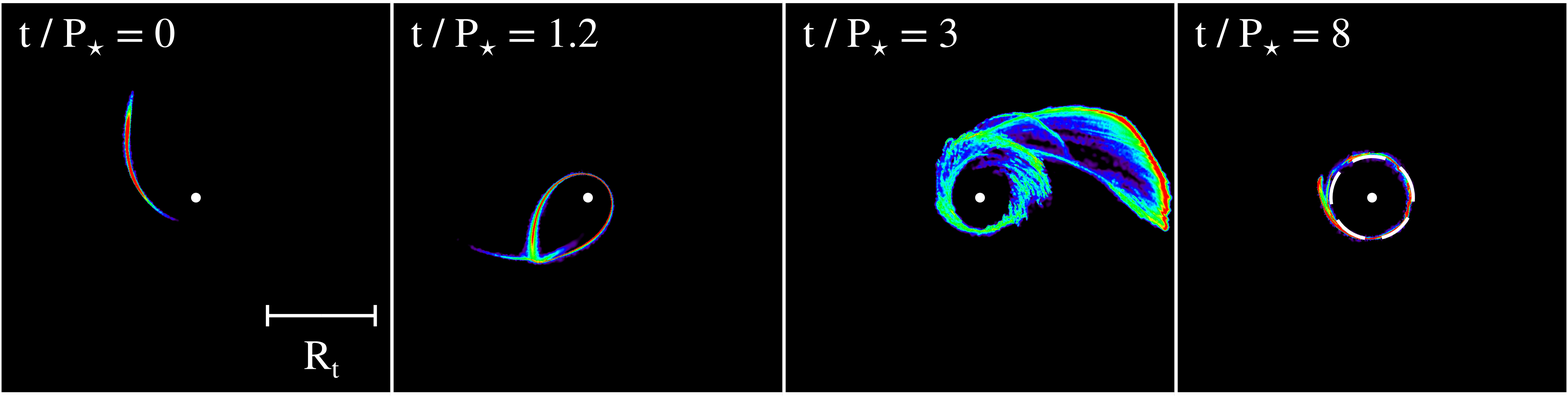}
\epsfig{width=\textwidth, file=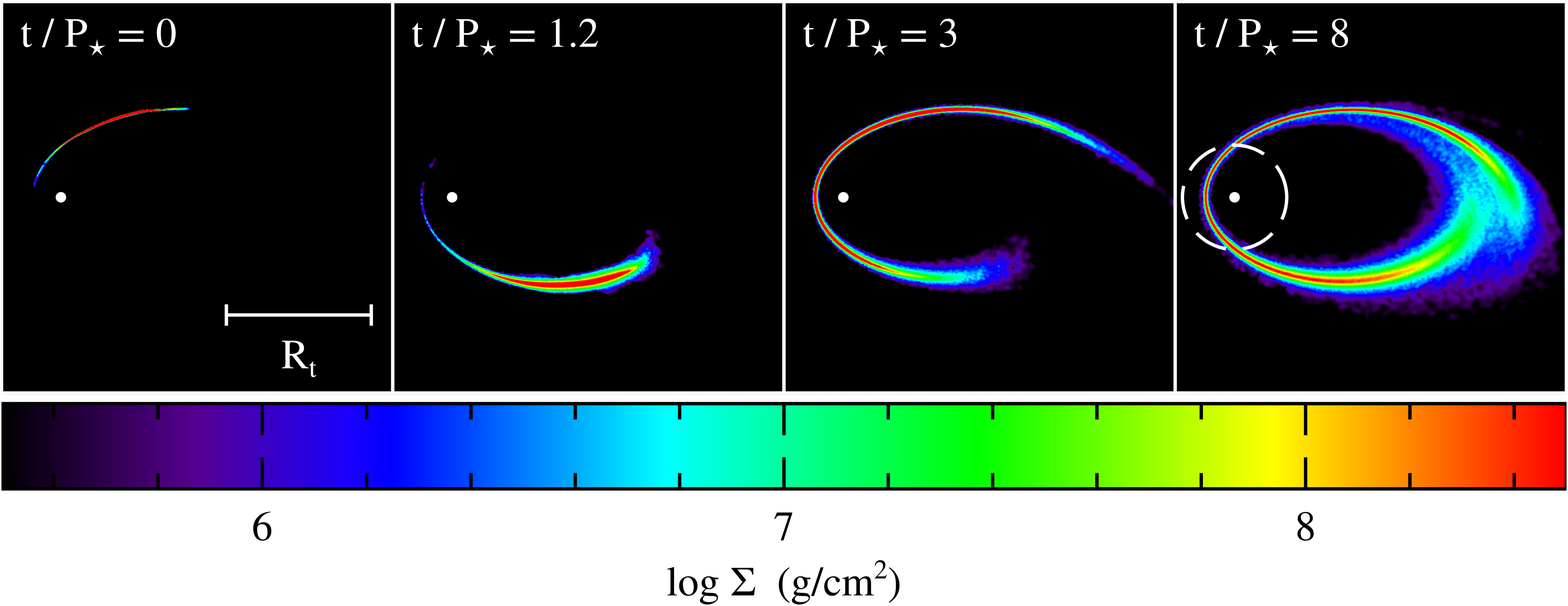}
\caption{Snapshots of the fallback of the debris at different times $t/ \pstar = 0, 1.2, 3$ and 8 for models RI5e.8 (upper panel) and KI5e.8 (lower panel). For these models, the period of the star $\pstar$ is $2.8 \, \rm h$. The colours correspond to the column density $\Sigma$ of the gas whose value is indicated on the colour bar. The white point represents the black hole. The dashed white circle on the last snapshot represents the circularization radius given by equations \eqref{rck} and \eqref{rcr} for the Keplerian and relativistic potentials respectively.}
\label{fig2}
\end{figure*}

For the disruption phase, the star is initially placed at the apocentre of its orbit, a distance $\ra=(1+e)\rp/(1-e)$ from the black hole. Its initial velocity $\va$ is determined by conservation of specific orbital energy and angular momentum between $\ra$ and $\rp$. In the Keplerian and relativistic potentials, the specific orbital energy is given by
\be
\varepsilon^{\rm K}=\frac{1}{2}\left(v^2_{\rm r} + v^2_{\rm t}\right)-\frac{G\mh}{R},
\ee
\be
\label{energyr}
\varepsilon^{\rm R}=\frac{1}{2}\left[\left(\frac{R}{R-2\rg}\right)^2 v^2_{\rm r} +\left(\frac{R}{R-2\rg}\right) v^2_{\rm t}\right]-\frac{G\mh}{R},
\ee
and the specific angular momentum by
\be
\label{angmomk}
l^{\rm K}=R v_{\rm t},
\ee
\be
\label{angmomr}
l^{\rm R}=\frac{R v_{\rm t}}{R-2\rg},
\ee
respectively.
The initial velocity of the star at apocentre is therefore
\begin{eqnarray}
\label{vak}
v^{\rm K}_{\rm a}  &=& \left(\frac{G \mh}{\ra}\right)^{1/2}\left(\frac{2\rp}{\ra + \rp}\right)^{1/2} \\ &=& \left(\frac{G \mh}{\ra}\right)^{1/2}\left(1-e\right)^{1/2},
\end{eqnarray}
\be
\label{var}
v^{\rm R}_{\rm a}=\left(\frac{G \mh}{\ra}\right)^{1/2}\frac{2^{1/2}\rp (\ra-2\rg)}{\ra((\ra+\rp)(\rp-2\rg)+2\rg\rp^2)^{1/2}},
\ee
depending on the potential.

In order to demonstrate the correct implementation of the relativistic potential into the SPH codes, we anticipate the next section and analyse the motion of the gas in the simulation associated to model RI5e.8. In this model, the relativistic potential is used and the orbit of the star has a penetration factor $\beta=5$ and an eccentricity $e=0.8$. Fig. \ref{fig1} shows the trajectory of the centre of mass of the gas in the simulation corresponding to model RI5e.8 compared to that of a test particle on the same orbit in the relativistic potential and in the Schwarzschild metric. The trajectory of the test particle in the relativistic potential and in the Schwarzschild metric is followed for 4 orbits. That of the centre of mass of the gas is followed during both the disruption of the star and the fallback of the debris until $t/\pstar = 8$. The trajectory of the test particle is the same in the relativistic potential and in the Schwarzchild metric, confirming that apsidal precession is treated accurately in the relativistic potential as found by \citet{tejeda2013}. During the disruption and the beginning of the fallback phase, the centre of mass of the gas and the test particle have the same trajectory, validating the implementation of the potential into the SPH codes. After a few pericentre passages, the two trajectories start to differ as the hydrodynamical effects on the debris become dominant. This evolution will be investigated in detail in the next section.

The accretion onto the black hole is modelled by an accretion radius fixed at the innermost stable circular orbit of $6 \rg$ for a non-rotating black hole. Particles entering this radius are removed from the simulations.

The simulation of the fallback phase is performed for two different equations of state (EOS) for the gas: locally isothermal and adiabatic. For the locally isothermal EOS, each SPH particle keeps their initial specific thermal energies. Physically, these two EOS represent the two extreme cases for the rate at which an excess of thermal energy is radiated away from the gas. For the locally isothermal one, every increase of thermal energy is instantaneously radiated away while for the adiabatic one, none of this energy is radiated.

Four parameters are therefore considered to simulate the fallback of the debris: the potential, the EOS, the penetration factor $\beta$ and the eccentricity $e$. The values of these parameters for the eight models investigated in this paper are shown in table \ref{param}.

\section{Results}
\label{results}

In this section, we present the results of the simulations for the fallback phase\footnote{Movies of the simulations presented in this paper are available at \url{http://home.strw.leidenuniv.nl/~bonnerot/research.html}.}. The time is scaled by the period of the star $\pstar$ with the starting point ($t/\pstar=0$) being when the most bound debris of the stream come back to pericentre. The circularization process is investigated through the evolution of both the position and the specific orbital energy of the debris. They will be respectively compared to the so called circularization radius and specific circularization energy.

\begin{table}
 \centering
  \caption{Name and parameters of the different models.}
  \begin{tabular}{@{}llrrrrlrlr@{}}
  \hline
   Model & Potential & EOS & $\beta$ & e  \\
 \hline
   RI5e.8 & Relativistic & Locally isothermal & 5 & 0.8   \\
   KI5e.8 & Keplerian & Locally isothermal & 5 & 0.8 \\
   RA5e.8 & Relativistic & Adiabatic & 5 & 0.8 \\
   KA5e.8 & Keplerian & Adiabatic & 5 & 0.8 \\
   RI1e.8 & Relativistic & Locally isothermal & 1 & 0.8 \\
   RA1e.8 & Relativistic & Adiabatic & 1 & 0.8 \\
   RI5e.95 & Relativistic & Locally isothermal & 5 & 0.95   \\
   RA5e.95 & Relativistic & Adiabatic & 5 & 0.95   \\
\hline
\end{tabular}
\label{param}
\end{table}

These two quantities are defined as follows. Before the disruption, the distribution of angular momentum of the gas inside the star is sharply peaked around that of the star. As this distribution is not significantly affected by the disruption, the debris still have similar angular momenta. Assuming that they then move from their initial eccentric orbits to circular orbits each of them conserving their angular momentum, these circular orbits will form at similar radii. It allows to define a characteristic circularization distance, called the circularization radius, obtained by equating the specific angular momentum of the star, given by equations \eqref{angmomk} and \eqref{angmomr} evaluated at apocentre, and the specific circular angular momentum given by
\be
\label{angmomck}
l^{\rm K}_{\rm c}=(G\mh R)^{1/2},
\ee
\be
\label{angmomcr}
l^{\rm R}_{\rm c}=\frac{(G\mh)^{1/2}R}{(R-3\rg)^{1/2}}.
\ee
This yields a circularization radius of
\be
\label{rck}
R^{\rm K}_{\rm circ}=\frac{\ra^2\va^2}{G\mh}=(1+e)\rp,
\ee
\be
\label{rcr}
R^{\rm R}_{\rm circ}=\frac{\ra^4\va^2+(\ra^4\va^2(-12 G \mh (\ra-2\rg)^2 \rg + \ra^4\va^2))^{1/2}}{2G\mh(\ra-2\rg)^2},
\ee
where $\va$ is obtained from equations \eqref{vak} or \eqref{var} depending on the potential. The associated specific orbital energy, called specific circularization energy, is equal to the specific circular orbital energy evaluated at the circularization radius. In the two potentials, it is given by
\be
\label{energyck}
\varepsilon^{\rm K}_{\rm circ}=-\frac{G\mh}{2\rcirc^{\rm K}}=-\frac{G\mh}{2(1+e)\rp},
\ee
\be
\label{energycr}
\varepsilon^{\rm R}_{\rm circ}=-\frac{G\mh}{2\rcirc^{\rm R}}\left(\frac{\rcirc^{\rm R}-4\rg}{\rcirc^{\rm R}-3\rg} \right).
\ee

\begin{figure}
\epsfig{width=0.47\textwidth, file=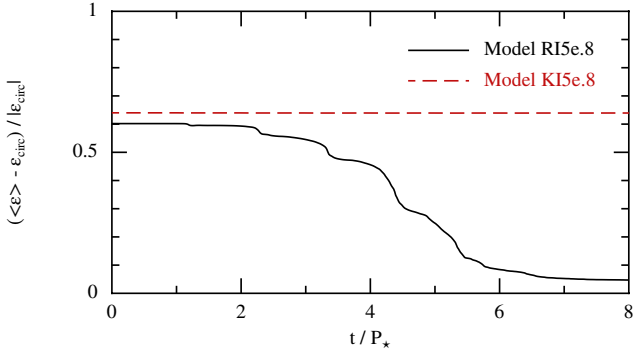}
\caption{Evolution of the average specific orbital energy of the debris for models RI5e.8 and KI5e.8. For these models, the period of the star $\pstar$ is $2.8 \, \rm h$. The average specific orbital energy is shown relative to the specific circularization energy given by equations \eqref{energyck} and \eqref{energycr} for the Keplerian and relativistic potentials respectively.}
\label{fig3}
\end{figure}

\begin{figure}
\epsfig{width=0.47\textwidth, file=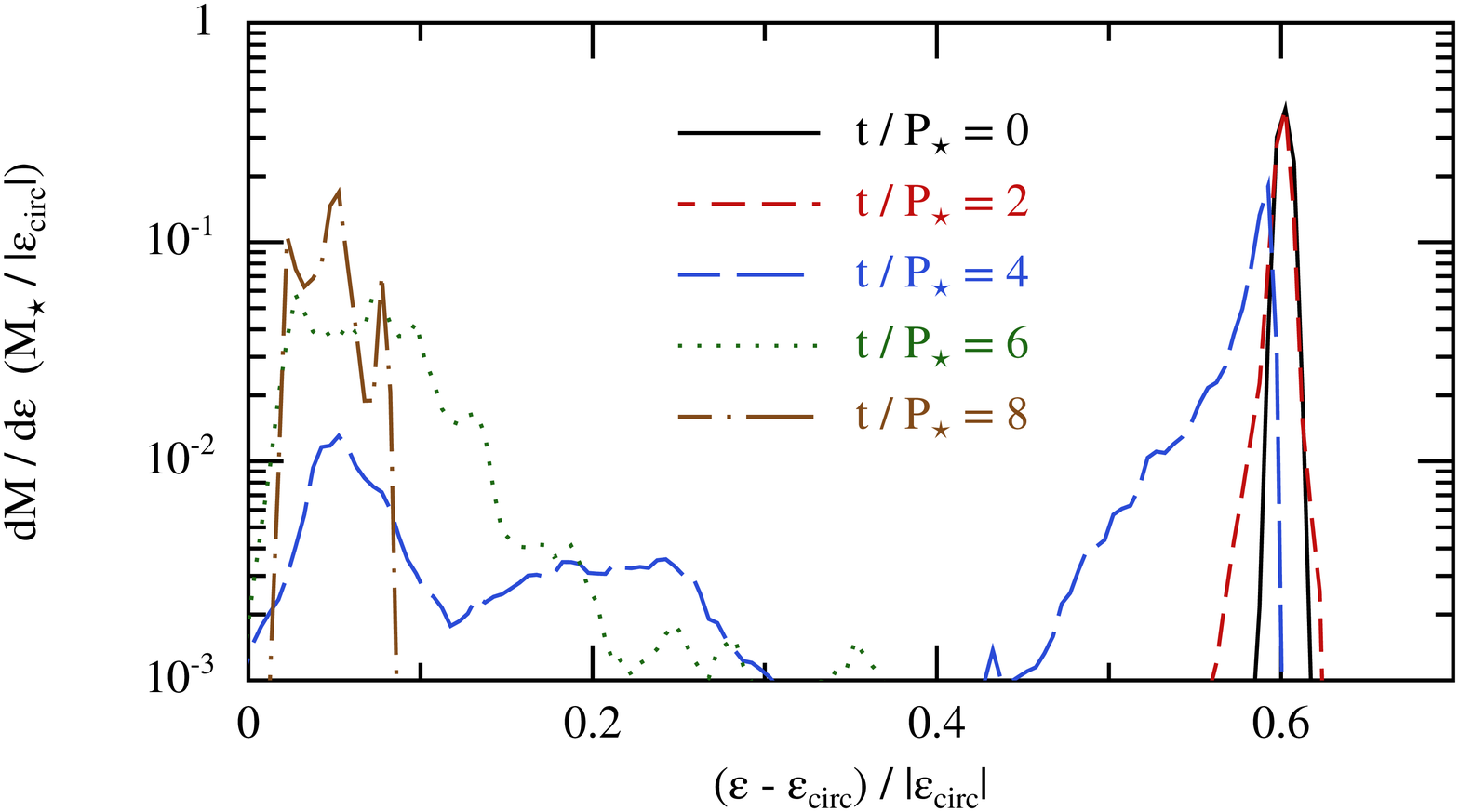}

\vspace{3mm}

\epsfig{width=0.478\textwidth, file=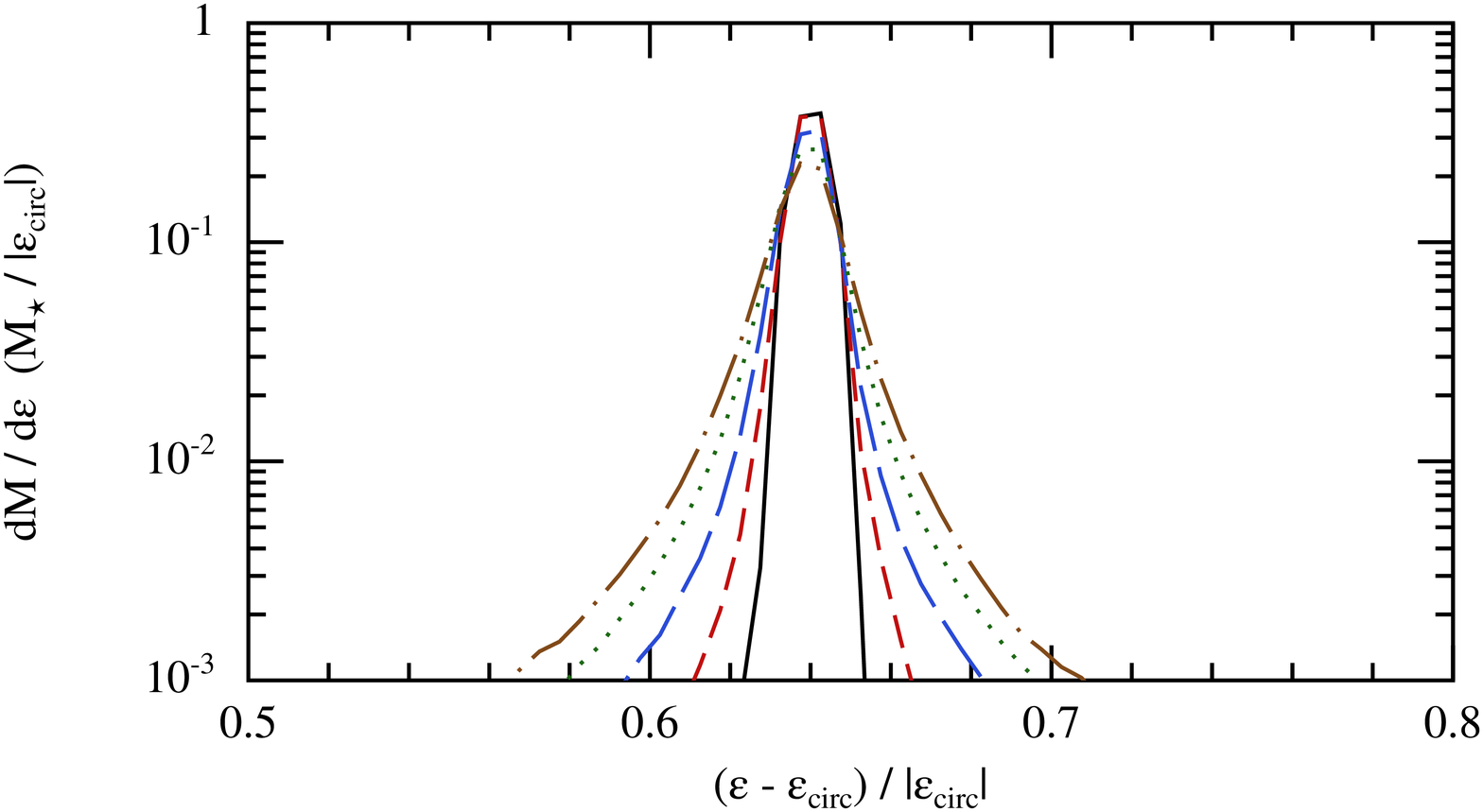}
\caption{Specific orbital energy distributions of the debris at different times $t/\pstar = 0, 2, 4, 6$ and 8 for model RI5e.8 (upper panel) and KI5e.8 (lower panel). For these models, the period of the star $\pstar$ is $2.8 \, \rm h$. The specific orbital energy is shown relative to the specific circularization energy given by equations \eqref{energyck} and \eqref{energycr} for the Keplerian and relativistic potentials respectively.}
\label{fig4}
\end{figure}

\subsection{Impact of relativistic precession}
\label{potential}

\begin{figure*}
\epsfig{width=\textwidth, file=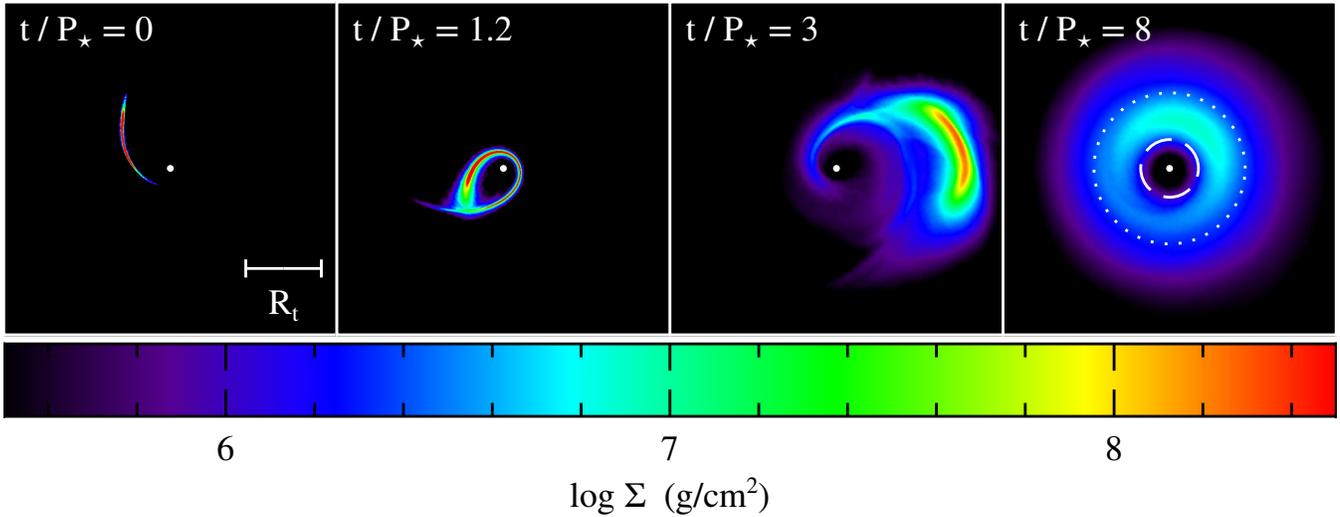}
\caption{Snapshots of the fallback of the debris at different times $t/\pstar = 0, 1.2, 3$ and 8 for model RA5e.8. For this model, the period of the star $\pstar$ is $2.8 \, \rm h$. The colours correspond to the column density $\Sigma$ of the gas whose value is indicated on the colour bar. The white point represents the black hole. The dashed white circle on the last snapshot represents the circularization radius given by equation \eqref{rcr} for the relativistic potential. The dotted white circle represents the semi-major axis of the star.}
\label{fig5}
\end{figure*}

As mentioned in the previous section, relativistic precession modifies the trajectory of the debris at pericentre. This effect is the strongest for the orbit with $\beta=5$ and  $e = 0.8$ where the precession angle reaches 89.7 degrees. This is because this orbit has the lowest pericentre distance $\rp=9.4\rg$ and, to a smaller extent, the lowest eccentricity. We investigate the role of relativistic precession for a star on this orbit by comparing models RI5e.8 and KI5e.8. In model RI5e.8, the relativistic potential is used and apsidal precession is taken into account. In model KI5e.8, the debris are forced to move on closed orbits by using a Keplerian potential. In both models, a locally isothermal EOS is adopted for the gas.

\begin{figure}
\epsfig{width=0.47\textwidth, file=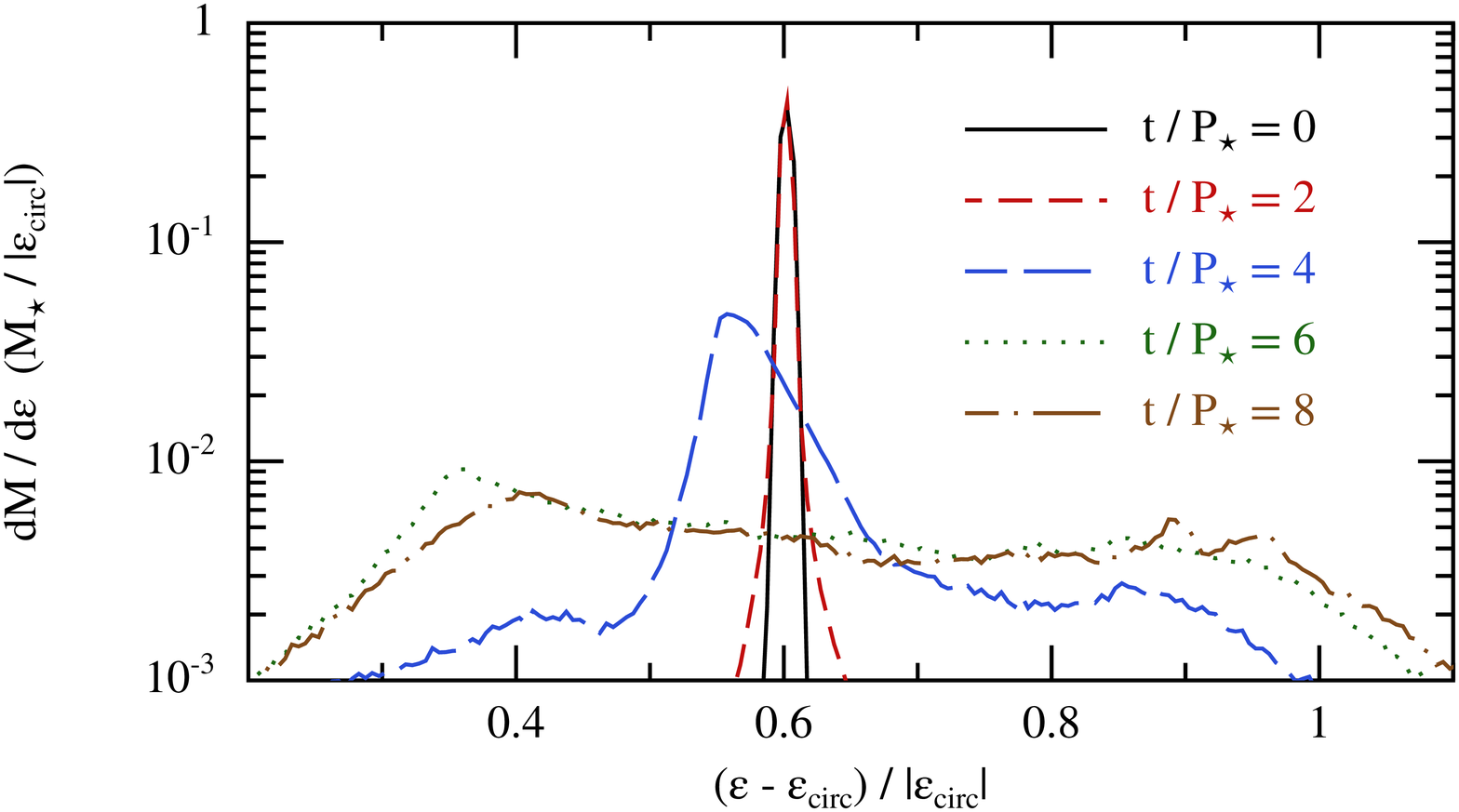}

\vspace{3mm}

\epsfig{width=0.47\textwidth, file=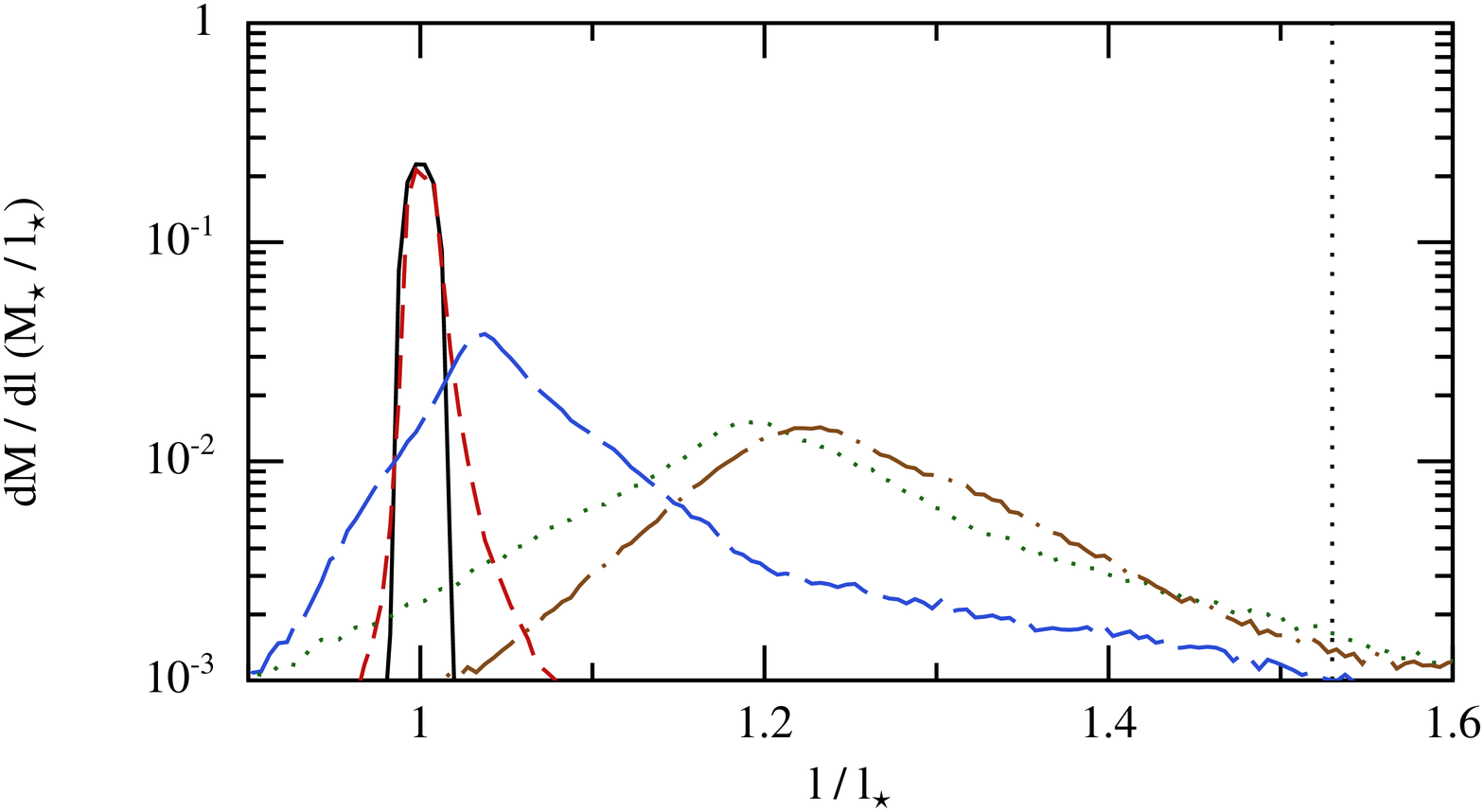}
\caption{Specific orbital energy distribution (upper panel) and specific angular momentum distribution (lower panel) of the debris at different times $t/\pstar = 0, 2, 4, 6$ and 8 for model RA5e.8. For this model, the period of the star $\pstar$ is $2.8 \, \rm h$. The specific orbital energy is shown relative to the specific circularization energy given by equation \eqref{energycr} for the relativistic potential. The specific angular momentum is shown relative to that of the star given by equation \eqref{angmomcr} evaluated at apocentre.}
\label{fig6}
\end{figure}

\begin{figure}
\epsfig{width=0.47\textwidth, file=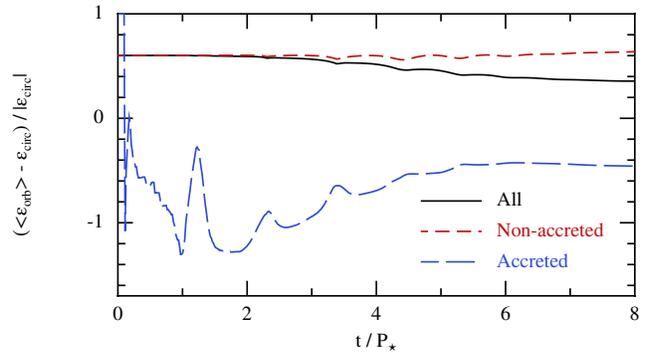}
\caption{Evolution of the average specific orbital energy of all the  debris, the non-accreted and accreted ones for model RA5e.8. For this model, the period of the star $\pstar$ is $2.8 \, \rm h$. The average specific orbital energy is shown relative to the specific circularization energy given by equation \eqref{energycr} for the relativistic potential.}
\label{fig7}
\end{figure}

We discuss model RI5e.8 first. The evolution of the debris can be seen in Fig. \ref{fig2} (upper panel), which shows snapshots of their fallback towards the black hole at different times $t/\pstar = 0, 1.2, 3$ and 8. The stream remains unaffected until $t/\pstar \simeq 1.2$, corresponding to its second pericentre passage. At this time, the stream self-crosses due to apsidal precession leading to the formation of shocks that convert kinetic energy into thermal energy. However, as a locally isothermal EOS is used, this excess thermal energy is instantaneously removed from the gas. The net effect of these shocks is therefore to reduce kinetic energy. This results in a decrease of the average specific orbital energy of the debris, whose evolution is shown in Fig. \ref{fig3} (solid black line). As self-crossings occur at each pericentre passage, this decrease continues for $t/\pstar \gtrsim 1.2$ through periodic phases. These phases of decrease also become progressively steeper as the self-crossings involve a larger fraction of the stream. Accordingly, the specific orbital energy distribution, shown in Fig. \ref{fig4} (upper panel), shifts towards lower energies. Owing to this orbital energy decrease, the debris progressively move from their initial eccentric orbits to circular orbits. At $t/\pstar \simeq 8$, their average specific orbital energy reaches a value similar to the specific circularization energy (Fig. \ref{fig3}, solid black line). By this time, they have settled into a thin and narrow ring of radius comparable to the circularization radius (Fig. \ref{fig2}, upper panel).

As can be seen from Figs. \ref{fig2} (upper panel) and \ref{fig3} (solid black line), the final specific orbital energy of the debris is somewhat larger than the specific circularization energy, which results in a ring forming slightly outside the circularization radius. These small discrepancies are due to redistribution of angular momentum between the debris during the shocks where a fraction of them (3 \% at $t/\pstar = 8$) loses enough angular momentum to be accreted onto the black hole. This causes an excess angular momentum shared by the remaining debris, that therefore settle into circular orbits at radii larger than the circularization radius. Notably, these discrepancies are reduced when the resolution of the simulations is increased, as will be shown in subsection \ref{convergence}.

We now discuss model KI5e.8 for which the evolution of the debris, shown in Fig. \ref{fig2} (lower panel), is very different. As there is no apsidal precession, the stream does not self-cross. Therefore, no kinetic energy is removed from the debris and their specific orbital energy remains constant (Fig. \ref{fig3}, dashed red line). Their specific orbital energy distribution also stays peaked around its initial value although it spreads somewhat due to tidal effects (Fig. \ref{fig4}, lower panel). Consequently, the debris do not move to circular orbits and settle instead at $t/\pstar \simeq 8$ into an elliptical strip centred around the initial orbit of the star (Fig. \ref{fig2}, lower panel).

The comparison of models RI5e.8 and KI5e.8 shows the fundamental role of apsidal precession in the circularization process. Our more accurate treatment of apsidal precession confirms the results of \citet{hayasaki2013}.

\subsection{Influence of the cooling efficiency}
\label{eos}

The models discussed in the previous subsection (RI5e.8 and KI5e.8) use a locally isothermal EOS for the gas. We now discuss models RA5e.8 and KA5e.8, that use an adiabatic EOS for the gas instead. The potential used is relativistic for model RA5e.8 and Keplerian for model KA5e.8. The star is on the same orbit as before with $\beta = 5$ and $e = 0.8$.

\begin{figure}
\epsfig{width=0.47\textwidth, file=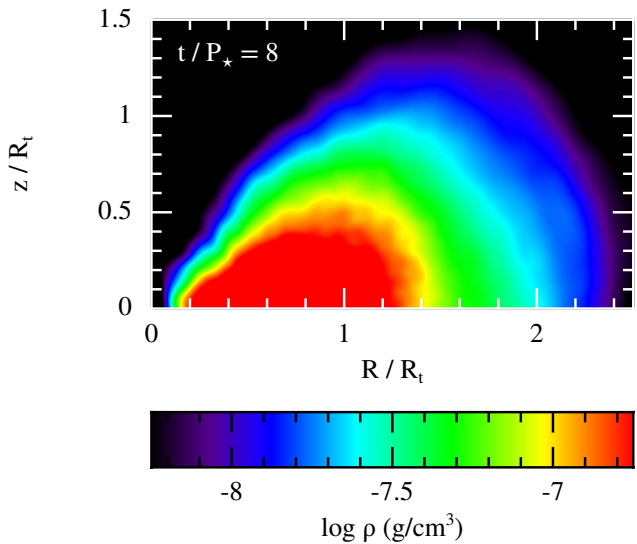}
\caption{Cross-section in the $R-z$ plane of the torus formed at $t/\pstar = 8$ for model RA5e.8. $z$ represents the height with respect to the midplane and $R$ the cylindrical radius. The distances are normalized by the tidal radius. The black hole is at the origin. The colours correspond to the density $\rho$ of the gas whose value is indicated on the colour bar.}
\label{fig8}
\end{figure}

\begin{figure}
\centering
\epsfig{width=0.47\textwidth, file=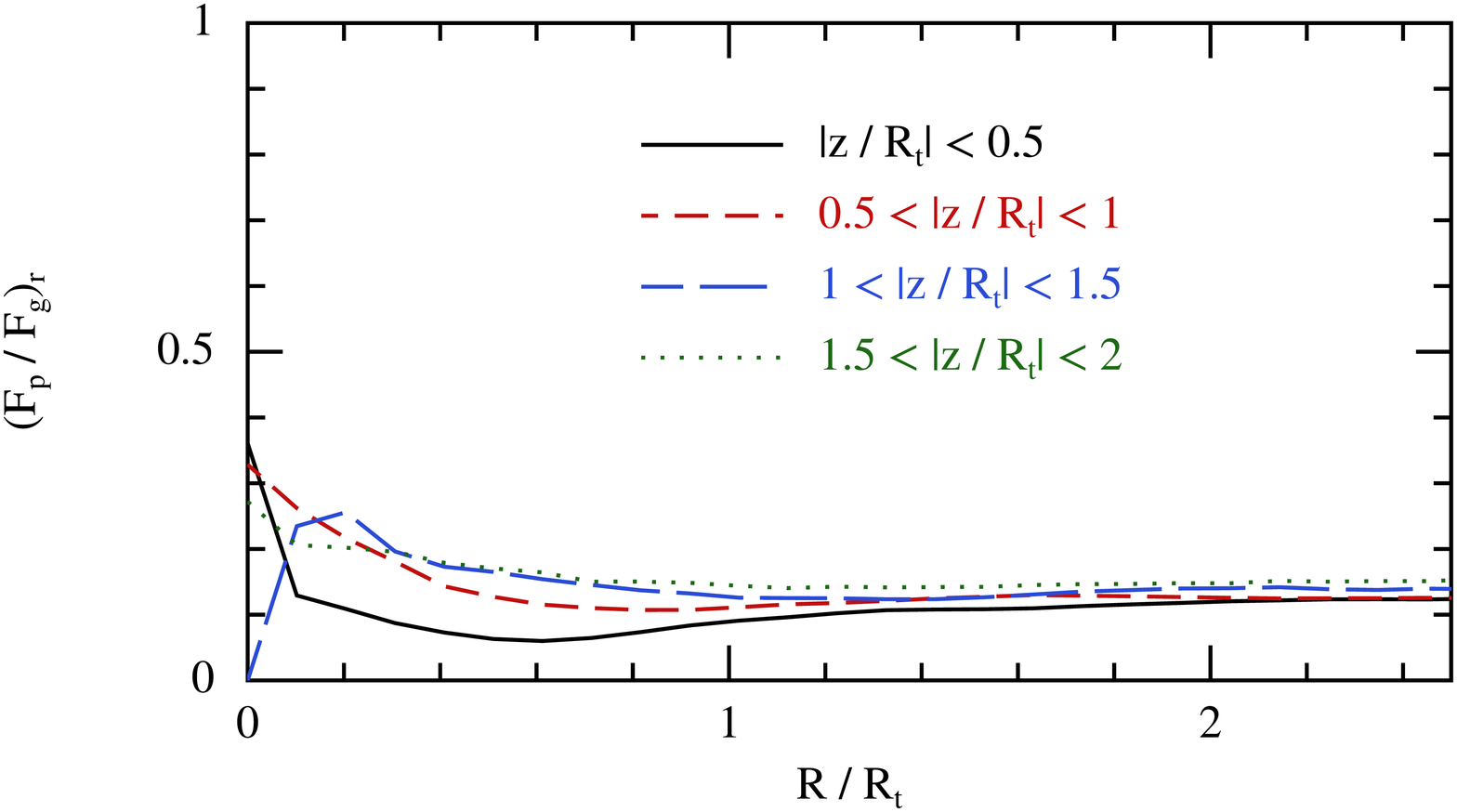}

\vspace{3mm}

\epsfig{width=0.47\textwidth, file=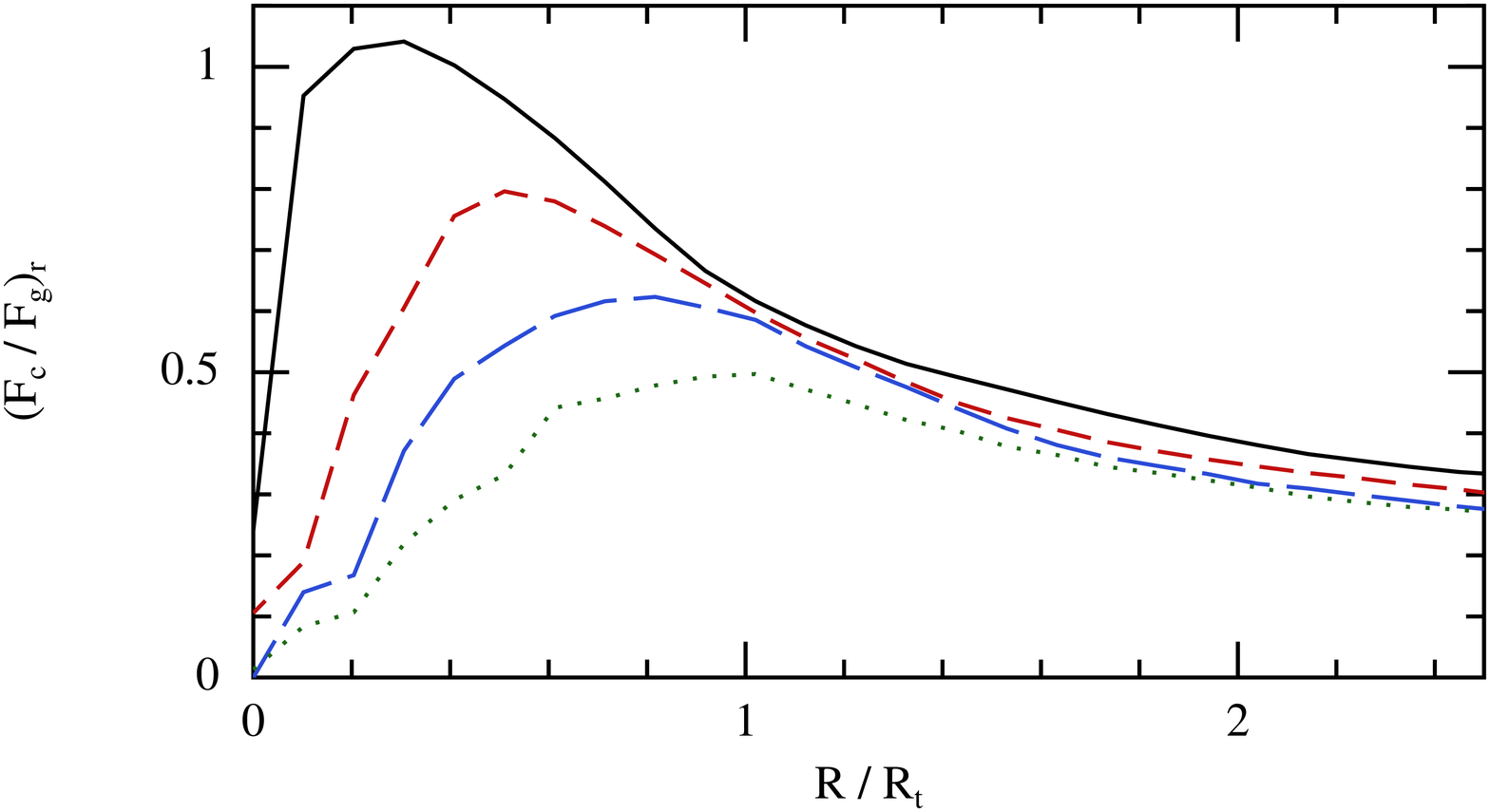}
\caption{Azimuthally-averaged ratios of the pressure (upper panel) and centrifugal (lower panel) forces to the gravitational force projected in the spherical radial direction as a function of the cylindrical radius $R$ at $t/\pstar = 8$ for model RA5e.8. These ratios are shown for different heights $z$ with respect to the midplane: $|z/\rt| < 0.5$, $0.5 < |z/\rt| < 1$, $1 < |z/\rt| < 1.5$ and $1.5 < |z/\rt| < 2$. The distances are normalized by the tidal radius.}
\label{fig9}
\end{figure}

We discuss model RA5e.8 first. The evolution of the debris as they fall back towards the black hole is shown in Fig. \ref{fig5} at different times $t/\pstar = 0, 1.2, 3$ and 8. The stream self-crosses at $t/\pstar \simeq 1.2$ and experiences shocks which convert kinetic energy into thermal energy. This early evolution is similar to model RI5e.8. However, the subsequent behaviour differs. As the EOS is adiabatic, the thermal energy produced by the shocks is not removed but kept in the debris. As more self-crossings occur, more thermal energy is injected into the stream which expands under the influence of thermal pressure. During the expansion, the ordered motion of the stream is suppressed and the debris depart from their initial eccentric orbits. This is achieved via a redistribution of their orbital parameters which can be seen in Fig. \ref{fig6} by a spread in the distributions of specific orbital energy (upper panel) and angular momentum (lower panel). The latter also presents a tail at large angular momenta that reaches, at $t/\pstar \simeq 8$, the specific circular angular momentum at the semi-major axis of the star, indicated in the lower panel of Fig. \ref{fig6} by a vertical black dotted line. Accordingly, the debris settle into a thick and extended torus with most of the gas located between the circularization radius and the semi-major axis of the star (Fig. \ref{fig5}). At this time, the majority of the debris (91\%) are still bound to the black hole as indicated by their negative specific orbital energies (Fig. \ref{fig6}, upper panel).

Remarkably, a significant fraction of the debris are ballistically accreted onto the black hole during the formation of the torus. The fraction of accreted gas grows roughly linearly to reach 26 \% at $t/\pstar \simeq 8$. Fig. \ref{fig7} shows the evolution of the average specific orbital energy of all the debris (solid black line), the non-accreted (dashed red line) and accreted (long dashed blue line) ones. When all the debris are considered, their orbital energy decreases due to the shocks where it is converted into thermal energy. However, it does not reach the circularization energy as part of the thermal energy is transferred back into kinetic energy during the expansion. The accreted debris are those which experienced the largest decrease of kinetic energy during the shocks as can be seen from their lower orbital energy. They also have the largest thermal energy which is therefore advected onto the black hole. The thermal energy of the torus is thus reduced due to both its expansion and the accretion of the hottest debris. Instead of decreasing, the orbital energy of the non-accreted debris stays constant. This is due to the accretion of debris with low orbital energies which results in an excess orbital energy shared among the non-accreted debris. As the accreted debris also have the lowest angular momenta, their accretion leads to an increase of the angular momentum of the remaining debris which results in a slight shift of the specific angular momentum distribution to larger values (Fig. \ref{fig6}, lower panel).

\begin{figure*}
\epsfig{width=\textwidth, file=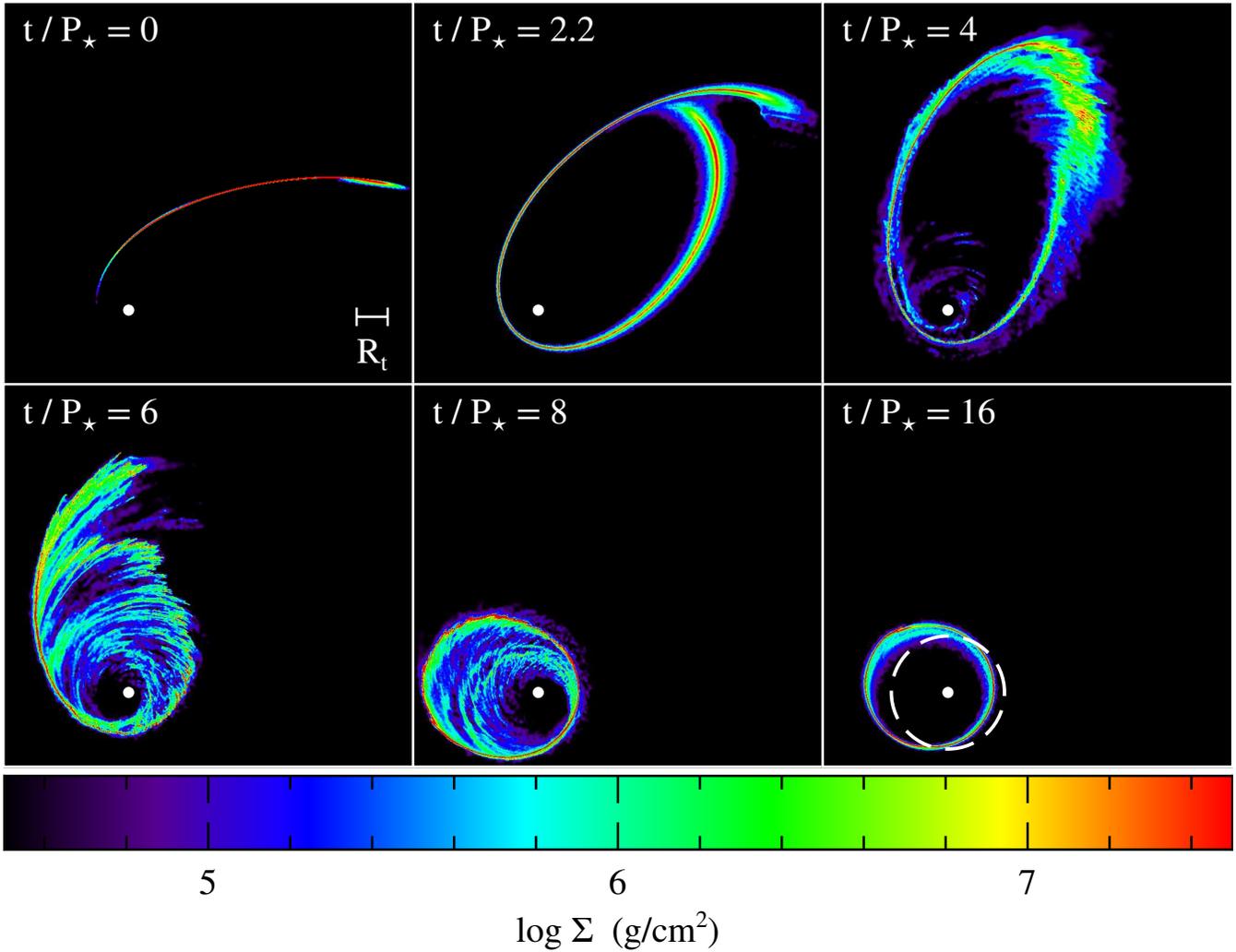}
\caption{Snapshots of the fallback of the debris at different times $t/\pstar = 0, 2.2, 4, 6, 8$ and 16 for models RI1e.8. For this model, the period of the star $\pstar$ is $31 \, \rm h$. The colours correspond to the column density $\Sigma$ of the gas whose value is indicated on the colour bar. The white point represents the black hole. The dashed white circle on the last snapshot represents the circularization radius given by equation \eqref{rcr} for the relativistic potential. Apsidal precession is weaker for this model than for model RI5e.8 owing to a larger pericentre, which causes the stream to self-cross further out from the black hole. At this location, relative velocities are lower, weakening the shocks. The debris therefore move slower to circular orbits.}
\label{fig10}
\end{figure*}

A more precise analysis of the torus formed at $t/\pstar = 8$ can be done by examining its internal configuration. Fig. \ref{fig8} shows a cross-section in the $R-z$ plane where $z$ represents the height with respect to the midplane and $R$ the cylindrical radius. It exhibits funnels around the rotation axis. It also contains a dense inner region at $z/\rt \lesssim 0.5$ and  $R/\rt \simeq 1$ that corresponds to the semi-major axis of the star. This dense region is surrounded by a more diffuse one that extends from close to the black hole to $R/\rt \simeq 2.5$ and $z/\rt \simeq 1.5$. To assess the internal equilibrium of the torus, we plot in Fig. \ref{fig9} azimuthally-averaged ratios of the pressure force (upper panel) and centrifugal force (lower panel) to the gravitational force projected in the spherical radial direction as a function of $R$ and for different intervals of $z$: $|z/\rt| < 0.5$, $0.5 < |z/\rt| < 1$, $1 < |z/\rt| < 1.5$ and $1.5 < |z/\rt| < 2$. The ratio of the pressure force to the gravitational force is roughly $(F_{\rm p}/F_{\rm g})_r \simeq 0.2$ in the entire torus implying that it is not hydrostatically supported. However, the ratio of the centrifugal force to the gravitational force is larger. For $|z/\rt| < 0.5$, it presents a maximum of $(F_{\rm c}/F_{\rm g})_r \simeq 1$ at $R/\rt \simeq 0.3$ and decreases at larger radii. The same dependence exists for regions further from the midplane but for lower maxima of $(F_{\rm c}/F_{\rm g})_r \simeq 0.5 - 0.8$. Therefore, this torus is mostly centrifugally supported against gravity with this support being stronger in its inner region. As they are bound to the black hole, the regions that are not supported against gravity will stop expanding and collapse at a later time. This collapse is likely to cause the formation shocks in the outer part of the torus which would increase hydrostatic support in this region.

For model KA5e.8, the stream does not experience apsidal precession. However, it is heated when it passes at pericentre. This is due to the formation of the pancake shock described in the Introduction. As a result, it expands roughly by a factor of 2 at each pericentre passage both in the radial and vertical directions. It causes the stream to self-cross at $t/\pstar \simeq 4$ that is after its fifth passage at pericentre. The subsequent evolution is similar to model RA5e.8. The debris settle into a thick and extended torus at $t/\pstar \simeq 10$. This channel of disc formation is similar to that found by \citet{ramirez-ruiz2009}. Therefore, relativistic apsidal precession is not the only factor that can lead to circular orbits of the debris, but it is the most efficient and operates regardless of the EOS used for the gas.

As can be seen by comparing models RI5e.8 and RA5.8, the cooling efficiency determines the structure of the disc formed during the circularization process. While a thin and narrow ring forms at the location of the circularization radius for an efficient cooling, an inefficient cooling leads instead to the formation of a thick and extended torus located between the circularization radius and the semi-major axis of the star.

\subsection{Dependence on the orbit of the star}

So far, the fallback of the debris has been investigated for a fixed orbit of the star with $\beta = 5$ and $e = 0.8$. We now consider two new orbits obtained by decreasing the penetration factor (subsection \ref{penetration}) and increasing the eccentricity (subsection \ref{eccentricity}). Throughout this section, the potential is fixed to relativistic. Models RI5e.8 and RAe.8, discussed above for the initial orbit and this potential, will be used as reference.

\subsubsection{Decreasing the penetration factor}
\label{penetration}

We start by decreasing the penetration factor to $\beta = 1$ keeping the eccentricity to its initial value $e = 0.8$. This decrease of the penetration factor corresponds to an increase of the pericentre distance from $\rp=9.4\rg$ to $\rp=47\rg$. We investigate the fallback of the debris for a star on this new orbit by discussing models RI1e.8 and RA1e.8. A locally isothermal EOS is used for model RI1e.8 while an adiabatic EOS is used for model RA1e.8. The relativistic potential is used in both models.

We discuss model RI1e.8 first. Snapshots of the evolution of the debris are shown in Fig. \ref{fig10} at different times $t/\pstar = 0, 2.2, 4,6,8$ and 16. The first self-crossing of the stream occurs at $t/\pstar = 2.2$ that is at its third pericentre passage. It causes the formation of shocks that reduce the kinetic energy of the debris as a locally isothermal EOS is used. However, as the stream passes further from the black hole, relativistic precession is weaker than for model RI5e.8 with a precession angle decreasing from 89.7 to 13.5 degrees. Consequently, the shocks happen further from the black hole and involve parts of the stream with lower relative velocities. For this reason, they are less efficient at removing kinetic energy from the debris. This results in a slower and more gradual decrease of their average specific orbital energy, as seen from Fig. \ref{fig11} that shows its evolution for model RI5e.8 (solid black line) and RI1e.8 (dashed red line). At $t/\pstar \simeq 16$, the average specific orbital energy of the debris stabilizes at a value similar to the circularization energy (Fig. \ref{fig11}, dashed red line) and the debris settle into a thin and narrow ring of radius comparable to the circularization radius (Fig. \ref{fig10}). As for model RI5e.8, the origin of the small discrepancies with the specific circularization energy and the circularization radius are due to the accretion onto the black hole of a fraction of the debris (3 \% at $t/\pstar = 16$) with low angular momenta.

For model RA1e.8, the evolution of the debris is very similar to model RA5.8. The shocks produced by the self-crossings of the stream leads to its expansion. The debris settle into a torus with most of them located between the circularization radius and the semi-major axis of the star. The only effect of modifying the orbit is to change the value of these two limiting radii.

\subsubsection{Increasing the eccentricity}
\label{eccentricity}

We now keep the penetration factor to its initial value $\beta=5$ while increasing the eccentricity to $e=0.95$. The fallback of the debris is investigated for this new orbit by discussing models RI5e.95 and RA5e.95 which use respectively a locally isothermal and an adiabatic EOS. The relativistic potential is adopted for both of them.

\begin{figure}
\epsfig{width=0.47\textwidth, file=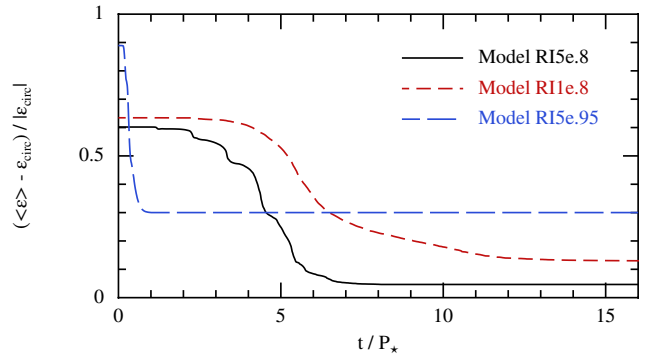}
\caption{Evolution of the average specific orbital energy of the debris for models RI5e.8, RI1e.8 and RI5e.95. For these models, the periods of the star $\pstar$ are respectively $2.8 \, \rm h$, $31 \, \rm h$ and $22 \, \rm h$. The average specific orbital energy is shown relative to the specific circularization energy given by equation \eqref{energycr} for the relativistic potential.}
\label{fig11}
\end{figure}

\begin{figure*}
\epsfig{width=\textwidth, file=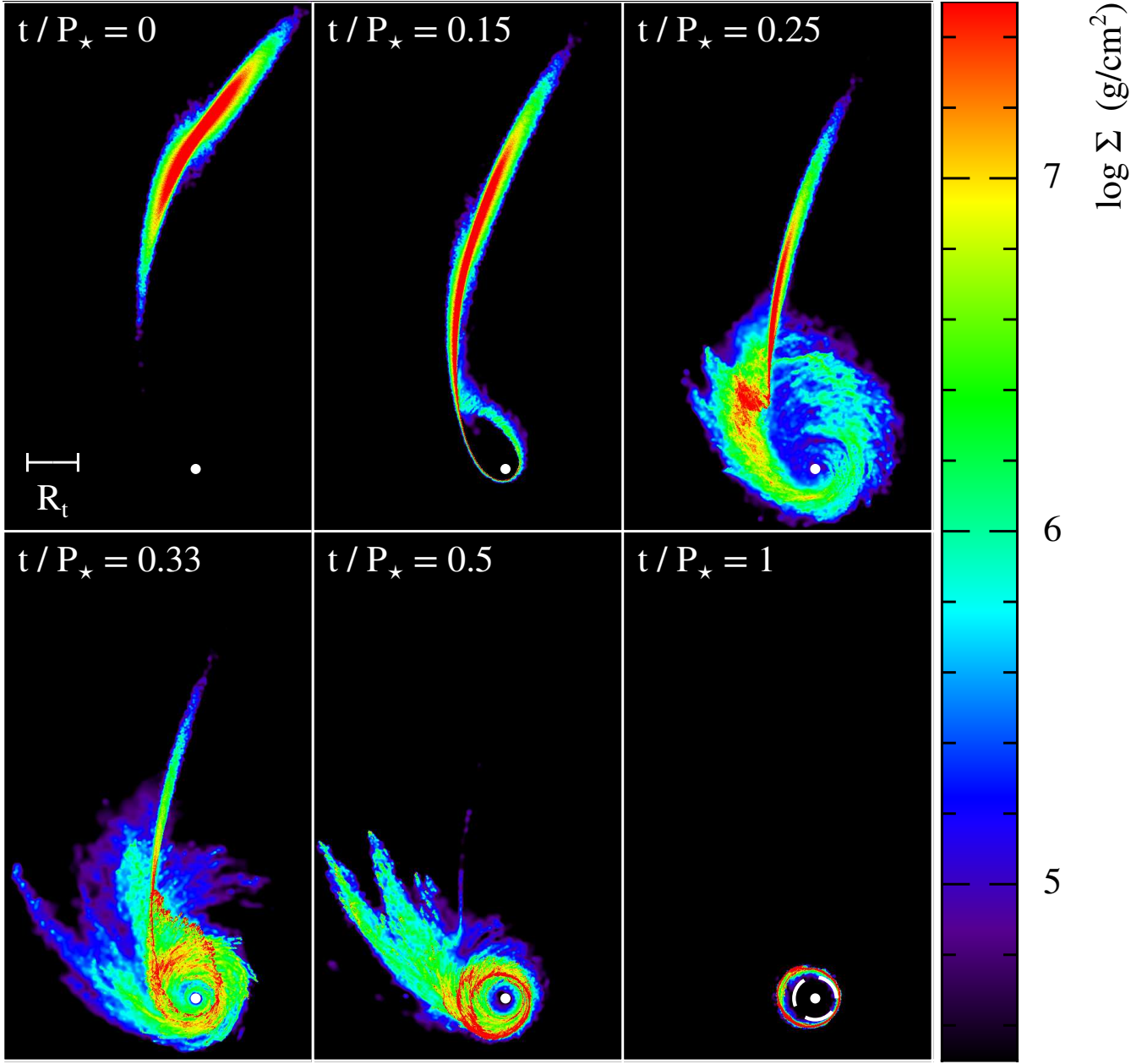}
\caption{Snapshots of the fallback of the debris at different times $t/\pstar = 0, 0.15, 0.5$ and 1 for models RI5e.95. For this model, the period of the star $\pstar$ is $22 \, \rm h$. The colours correspond to the column density $\Sigma$ of the gas whose value is indicated on the colour bar. The white point represents the black hole. The dashed white circle on the last snapshot represents the circularization radius given by equation \eqref{rcr} for the relativistic potential. The stream is longer for this model than for model RI5e.8 as the eccentricity is larger. As a result, the self-crossings involve a larger fraction of the debris, causing them to move faster to circular orbits.}
\label{fig12}
\end{figure*}

We discuss first model RI5e.95. Fig. \ref{fig12} shows snapshots of the fallback of the debris at different times $t/\pstar = 0, 0.15, 0.25, 0.33, 0.5$ and 1. The stream first crosses itself at $t/\pstar \simeq 0.15$ just after the first passage of its leading part at pericentre. When this occurs, most of the debris are still falling back towards the black hole away from the self-crossing point. This is different from model RI5e.8, for which this self-crossing happens later and when most of the debris are already beyond the self-crossing point (see Fig. \ref{fig2}). The reason for these differences is that the stream is longer for model RI5e.95 than for model RI5e.8 owing to the larger eccentricity. After the first self-crossing, the debris located beyond the self-crossing point are expelled leaving the rest of the stream free to move around the black hole. This induces a second self-crossing at $t/\pstar \simeq 0.33$. These self-crossings create shocks that remove kinetic energy from the debris as a locally isothermal EOS is used. As shown in Fig. \ref{fig11} (blue long dashed line), the average specific orbital energy of the debris therefore decreases for $t/\pstar \gtrsim 0.15$. This decrease occurs faster than for model RI5e.8 as the self-crossings involve larger fractions of the debris. At $t/\pstar \simeq 1$, their average specific orbital energy settles at a value similar to the circularization energy (Fig. \ref{fig11}, blue long dashed line) and they form a thin and narrow ring of gas located at a distance comparable to the circularization radius (Fig. \ref{fig12}). As previously, the discrepancies with the specific circularization energy and the circularization radius are due to the accretion of a fraction of the debris (4 \% at $t/\pstar = 1$) with low angular momenta. These discrepancies are larger for this model than for models RI5e.8 and RI1e.8 because more debris are accreted and these debris have lower angular momenta relative to that of the star.

For model RA5e.95, the debris evolve very similarly to model RA5.8. They expand due to shocks produced by the self-crossings of the stream. They settle into a thick and extended torus in which most of the debris are located between the circularization radius and the semi-major axis of the star. Only the value of these two limiting radii differs for the new orbit.

\subsection{Convergence of the results}
\label{convergence}

The simulations of the fallback phase has been performed for three different resolutions corresponding to about 100K, 500K and 1300K particles. The two larger resolutions have been obtained from the lowest one by using the particle splitting technique \citep{kitsionas2003} on the initial condition of the fallback phase. Each particle is split into 5 or 13 additional ones with one at the position of the initial particle and the others distributed on a tetrahedron and on a face-centred cubic structure respectively. In both cases, the distance between each additional particles is fixed to 1.5 times the smoothing length of the initial one. In addition, the particles added outside the volume defined by the initial ones have been removed. The fraction of particles removed in this way is always less than 10\%. The mass of the additional particles has then been decreased in order to keep the total mass of the stream constant.

Minor differences have been noticed when the resolution is increased. They are common to the different orbits considered. For a locally isothermal EOS, the time at which the average specific orbital energy of the debris settles is either advanced or delayed by $\Delta t \simeq \pstar$ between the 100K and the 500K simulations. However, this difference becomes negligible between the 500K and 1300K simulations. Both for a locally isothermal and an adiabatic EOS, increasing the resolution also results in a smaller fraction of debris with low angular momenta being accreted onto the black hole. For the isothermal EOS, this causes the average specific orbital energy of the debris to settle at a value closer to the specific circularization energy. The fraction of debris accreted during the disc formation decreases to less than 5\% in this case. We therefore attribute it to resolution. For the adiabatic EOS, the torus turns out to be slightly more compact and centrally condensed. However, its internal structure remains the same. The fraction of accreted debris also decreases in this case but remains larger than 25\%, and is likely to be physical. The simulations presented in this paper have been performed for 500K particles as both the final configuration of the debris and the time needed to reach it are unchanged above this resolution. Overall, we conclude that the behaviour described in the previous subsections is robust with respect to resolution.\\

\section{Discussion and conclusion}
\label{discussion}

By means of SPH simulations, we investigated the circularization process for tidal disruptions of stars on bound orbits by a non-rotating black hole. The formation of an accretion disc from the debris is mostly driven by relativistic apsidal precession that causes the stream to self-cross. If cooling is inefficient, this self-crossing is also partially caused by an expansion of the stream, when it passes at pericentre. In addition, we showed that the structure of the disc depends on the cooling efficiency of the gas by considering two extreme cases. For an efficient cooling, the debris form a thin and narrow ring of gas. For an inefficient cooling, they settle into a thick and extended torus, that at the end of our simulation is still mostly centrifugally supported against gravity. We also demonstrated the existence of different regimes of circularization for different orbits of the star. The circularization timescale $\tcirc$ varies from $\sim \pstar$ for the largest eccentricity to $\sim 10 \pstar$ for the lowest penetration factor considered (see Fig. \ref{fig11}). In physical units, it corresponds to tens to hundreds of hours.

The circularization process in TDEs has been the focus of several other recent works. \citet{shiokawa2015} simulated the tidal disruption of a white dwarf on a parabolic orbit around a $500 \msun$ black hole using a general relativistic simulation while \citet{hayasaki2015} considered the case of bound orbits around a $10^6 \msun$ black hole. The latter use a simulation setup similar to ours but our treatment of apsidal precession is more accurate and we consider a larger range of orbits. Both studies noticed the influence of the cooling efficiency on the disc structure. In addition, \citet{hayasaki2015} investigated the effect of the black hole spin. As anticipated by \citet{dai2013}, they found that nodal precession can prevent the self-crossing of the stream, delaying circularization. Using Monte-Carlo calculations, \citet{guillochon2015} found that this delay is of one year on average for a star on a parabolic orbit.

\subsection{Thermal energy radiation}
\label{radiation}

During the circularization process, thermal energy is injected into the debris at the expense of their kinetic energy. If they cool efficiently, this thermal energy increase can be estimated by subtracting the final circularization energy to the initial orbital energy of the star. It amounts to $\Delta U = \mstar (\varepsilon^{\rm R} - \varepsilon^{\rm R}_{\rm circ}) \simeq \rm{few} \,\, 10^{51} - 10^{52} \, \erg$ for the orbits considered, according to equations \eqref{energyr} and \eqref{energycr}. For larger $\beta$, $\Delta U$ is larger because the circularization energy is lower. Assuming that this excess thermal energy is radiated during the disc formation, it gives rise to a flare of luminosity $\lcirc = \Delta U / \tcirc \simeq \rm{few} \,\, 10 - 10^3 \, \ledd$, increasing with $\beta$, as $\tcirc$ is shorter and $\Delta U$ is larger.

As described in subsection \ref{eos}, if the excess thermal energy is not radiated instantaneously, it is partially transferred back into kinetic energy or advected onto the black hole. As a result, the remaining thermal energy of the torus is $\sim 10\%$ of the above $\Delta U$. The associated heating rate is therefore $\Delta U / \tcirc \simeq \rm{few} \,\, 1 - 10^2 \, \ledd$. Note that at the end of our simulation, the torus has not yet settled into an equilibrium configuration as we observe outflowing material.

\subsection{Viscous evolution}
\label{viscous}

Debris on circular orbits are subject to viscous effects which drive their accretion onto the black hole. Viscosity is not explicitly included in the simulations presented above. Nevertheless, its influence can be estimated a posteriori by computing the viscous timescale $\tvisc$. In our simulations, disc formation is progressive. A fraction of the debris already forms a disc-like structure before complete circularization. An example of such a structure can be seen in Fig. \ref{fig12} at $t/\pstar=0.33$. We consider that viscosity can start acting on these debris at $t<\tcirc$ which may lead to their accretion before complete circularization if $\tvisc<\tcirc$. For simplicity, $\tvisc$ is computed for the final configuration of the disc although it is used to evaluate viscous effects as the debris proceed towards this configuration. If the debris cool efficiently, they settle into a thin ring around the circularization radius. At this distance from the black hole,
\be
\frac{\tvisc}{\pstar}= 3 \times 10^3 \left(\frac{\alpha}{0.1} \right)^{-1} \left(\frac{H/R}{10^{-2}} \right)^{-2} \left(\frac{1-e^2}{0.36} \right)^{3/2},
\label{tviscisot}
\ee
where $\alpha$ is the viscosity parameter \citep{shakura1973} and $H/R$ is the aspect ratio of the disc. The circularization radius is obtained from equation \eqref{rck} which is at the origin of the $(1-e^2)^{3/2}$ factor. For all orbits considered, $\tvisc > \tcirc$, largely independently of the values of $\alpha$ and $H/R$. Therefore, viscosity should not have a significant effect during circularization and most of the accretion will occur once the ring is formed. At the end of circularization, the viscous timescale, given by equation \eqref{tviscisot}, corresponds to an accretion rate $\mdot = \mstar/\tvisc \simeq \rm{few} \,\, 1 - 100 \, \mdotedd$ depending on the orbit, $\mdotedd$ being the Eddington accretion rate for a radiative efficiency of 10\%. We therefore speculate that, due to this super-Eddington accretion, the ring will subsequently evolve into a thicker structure under the influence of radiation pressure.

If the gas cools inefficiently, it settles into a thick torus located between the circularization radius and the semi-major axis of the star. At the semi-major axis of the star,
\be
\frac{\tvisc}{\pstar}= 2 \, \left(\frac{\alpha}{0.1} \right)^{-1} \left(\frac{H/R}{1} \right)^{-2},
\label{tviscisos}
\ee
which does not depend explicitly on the orbit of the star as $\pstar$ cancels out. For the above values of $\alpha$ and $H/R$, $\tvisc \lesssim \tcirc$ which is sensitive to the values chosen for these parameters. However, when evaluated at the circularization radius, $\tvisc$ is lower by a factor $(1-e^2)^{3/2} \simeq \rm{few} \,\, 10^{-2} -10^{-1}$ depending on the orbit. The precise value of the viscous timescale thus depends on the mass distribution within the torus, being shorter if more mass is close to the circularization radius. We therefore conclude that viscosity may affect the evolution of the debris during the circularization process causing some of them to be accreted. During its subsequent evolution, the torus will keep accreting matter with an accretion rate $\mdot = \mstar/\tvisc \gtrsim  10^4 \, \mdotedd$ for the different orbits considered. For these highly super-Eddington accretion rates, the subsequent evolution of the torus is difficult to predict.

\subsection{Evaluation of the cooling efficiency}
\label{evalcool}

In the simulations presented in this paper, two extreme cooling efficiencies have been considered. The ability of the debris to cool can be estimated a posteriori by computing the diffusion timescale $\tdiff$ defined as the time that photons take to diffuse out of the surrounding gas. As most of the thermal energy is produced by the shocks occurring when the stream self-crosses, the diffusion timescale must be evaluated at the location of these shocks. Since we find that the gas is optically thick to electron scattering, it is given by $\tdiff=H_{\rm sh} \, \tau/c$ where  $\tau=\sigmat \rho_{\rm sh} H_{\rm sh} /\mp$ is the optical depth, $H_{\rm sh}$ and $\rho_{\rm sh}$ are the width and density of the gas at the location of the shocks. When an efficient cooling is assumed, the width of the stream remains $H_{\rm sh} \simeq \rstar$. To ensure self-consistency, the condition $\tdiff < \tcirc$ has to be satisfied, which translates into an upper limit on the density
\be
\rho_{\rm sh} < 8 \times 10^{-7} \gcm3 \, \left(\frac{\ncirc}{5} \right) \left(\frac{H_{\rm sh}}{\rstar} \right)^{-2} \left(\frac{\astar}{100 \rsun} \right)^{3/2},
\label{denscond}
\ee
where $\ncirc=\tcirc/\pstar$. Among the different orbits considered, this condition is satisfied only for that with $\beta=1$. In general, we estimate that increasing the eccentricity $e$ of the star tends to favour an efficient cooling of the debris. This is because it leads to a more extended and tenuous stream, decreasing $\rho_{\rm sh}$. Condition \eqref{denscond} is therefore more easily fulfilled. Instead, increasing the penetration factor $\beta$ may favour an inefficient cooling. It causes the stream to self-cross closer to the black hole. The debris are therefore located in a smaller volume which increases $\rho_{\rm sh}$. Furthermore, they circularize faster which decreases $\ncirc$. Condition \eqref{denscond} is thus more difficult to fulfill. However, these estimates are only approximate. The ability of radiation to escape depends on the precise location within the stream at which the thermal energy is deposited by the shocks. It also varies with the density distribution of the debris which is highly inhomogeneous at the time when most of the shocks occur. Furthermore, this radiation could also affect the structure of the disc through radiation pressure. A realistic treatment of the interaction between gas and radiation is therefore necessary to determine precisely the influence of cooling during the circularization process.

\subsection{Extrapolation to parabolic orbits}

As mentioned in the Introduction, TDEs typically involve stars on parabolic orbits. In this paper, we chose to simulate the disruption of stars on bound orbits instead. This choice is motivated by a lower computational cost which allows to explore a larger parameter space. However, the results can be extrapolated to get insight into the typical case of parabolic orbits.

The bound orbits considered in the simulations satisfy the condition $e < \ecrit = 1-(2/\beta)(\mh/\mstar)^{-1/3}$, so that all the debris are bound to the black hole. Furthermore, all the debris have periods similar to that of the star, $\pstar$. Instead, in the case of parabolic orbits, part of the debris become unbound from the black hole. Therefore, the tidal stream has a large range of periods between that of the most bound debris $\tmin$ and $+\infty$. This means that it features an elongated tail of debris which will continue to fall back towards pericentre long after the most bound ones have reached it. Due to apsidal precession, the leading part of the stream will inevitably collide with this tail after its first passage at pericentre. By means of point particle calculations in the Schwarzschild metric, we found that this prompt self-crossing occurs in general for an eccentricity $e \gtrsim 0.9$ largely independent on $\beta$. As discussed in subsection \ref{eccentricity}, it happens for model RI5e.95 where $e=0.95$ and $\beta=5$ (see Fig. \ref{fig12}). In this case, the self-crossing leads to circularization on a timescale $\tcirc \simeq \pstar$. Therefore, for a parabolic orbit with $\beta=5$, we expect a disc to form from the most bound debris due to this prompt self-crossing on a timescale $\tcirc \simeq \tmin$ where $\tmin$ replaces $\pstar$ as the period of the most bound debris. In this case, the newly-formed disc is only composed of the tip of the stream, which falls back within $\sim \tmin$. We expect the debris infalling later to rapidly circularize and join this disc. The timescale $\tcirc$ for the most bound debris to circularize is less clear for a parabolic orbit with $\beta=1$ as apsidal precession is weaker. For model RI1e.8 where $e=0.8$ and $\beta=1$, it is $\tcirc \simeq 10 \pstar$. However, the presence of a tail of debris would cause the self-crossing to affect the middle of the stream instead of its extremities, which likely makes the shocks more disruptive. On the other hand, the self-crossing happens further out from the black hole. This is because the apocentre of the most bound debris is about ten times larger for a parabolic orbit than for a bound one with $e=0.8$. Relative velocities are lower at this location, which likely weakens the shocks. Therefore, $\tcirc$ may remain $\sim 10 \, \tmin$ for a parabolic orbit with $\beta=1$. In this case, a significant mass of debris falls back within $\tcirc$, while the most bound ones circularize. Whether all this mass has circularized by $\tcirc$ is unclear.

Based on our estimates in subsection \ref{evalcool}, conditions for efficient cooling are more easily met for parabolic orbits than for elliptical ones. However, how efficient the cooling is also depends on the penetration factor, as large $\beta$ favours the formation of denser regions. In general, we expect the cooling efficiency to have the same effect on the disc structure as found in our simulations. If cooling is efficient throughout the evolution, a thin ring forms around the circularization radius. If it is inefficient, a thick torus forms located between the circularization radius and around the semi-major axis of the most bound debris. This allows us to extrapolate to the case of parabolic orbits the discussion in subsections \ref{radiation} and \ref{viscous}.

If cooling is efficient, the increase of thermal energy experienced by the debris during circularization is equal to the total change in orbital energy of the most bound debris. It amounts to $\Delta U = f \mstar (\varepsilon^{\rm R} - \varepsilon^{\rm R}_{\rm circ}) \simeq \rm{few} \,\, 10^{51} \, \erg$ according to equations \eqref{energyr} and \eqref{energycr}. For $\beta=1$, this estimate assumes that all the debris falling back within $\tcirc$ have circularized. The factor $f$ accounts for the fact that only a fraction of the debris reaches the black hole within $\tcirc$ in the parabolic case. Numerically, $f\simeq 0.2-0.4$ for $\tcirc = 1- 10 \, \tmin$ assuming a flat energy distribution. If this thermal energy is radiated during $\tcirc$, it leads to a luminosity $\lcirc=\Delta U/\tcirc \simeq  \rm{few} \,\, 1 - 10 \, \ledd$. Remarkably, it is comparable or higher than the peak luminosity in the soft X-ray band from the viscous accretion accretion through the disc \citep{lodato2011}. If the thermal energy is not immediately radiated but partly used to expand the disc or advected onto the black hole, the remaining thermal energy of the torus is $\sim 10 \%$ of the above $\Delta U$, a percentage extrapolated from the case of bound orbits. The heating rate is then $\Delta U/\tcirc \simeq  \rm{few} \,\, 0.1-1 \, \ledd$. This roughly agrees with the value found by \citet{shiokawa2015} scaling their results to our parameters for the black hole and the star \citep{piran2015}. 

The effect of viscosity can be estimated in the case of parabolic orbits by computing the viscous timescale $\tvisc$, obtained from equations \eqref{tviscisot} and \eqref{tviscisos} replacing $\pstar$ and $e$ by the period and eccentricity of the most bound debris $\tmin$ and $\ecrit$. As for bound orbits, we expect the disc to form progressively, with a fraction of the debris rapidly reaching a disc-like structure. For $\beta=1$, these circularized debris could already be present before most of the final disc mass reaches pericentre. On these debris, viscosity may start acting even before the disc completely settles. If cooling is inefficient, we find that $\tvisc \simeq \rm{few} \, \, \tmin$ at the semi-major axis of the most bound debris but decreases by a factor $(1-\ecrit^2)^{3/2} \simeq \rm{few} \,\, 10^{-3} - 10^{-4}$ at the circularization radius. The distribution of mass within the torus thus determines the value of the relevant viscous timescale which decreases if most of the debris are close to the circularization radius. In all cases, we expect $\tvisc < \tcirc$ since $\tcirc$ is estimated to be $\gtrsim \tmin$. We conclude that a significant fraction of the torus may already be accreted by the end of circularization. If cooling is efficient, $\tvisc \simeq \rm{few} \, \, 10-100 \, \tmin$ in the thin ring. In this case, $\tvisc > \tcirc$ which indicates that the ring is not significantly accreted during circularization. However, if the disc geometry changes when accretion starts, the viscous timescale could be different. In particular a thicker disc may result from a super-Eddington accretion rate, shortening the viscous timescale.

The fact that $t_{\rm visc} \ll t_{\rm min}$ has often been interpreted as evidence that the accretion rate onto the black hole traces directly the fallback rate of the debris. However, if most of the debris initially form a  disc without having the possibility of accreting, then the accretion rate onto the black hole will be driven by viscous processes rather than by infall, which has essentially finished once accretion starts \citep{cannizzo1990,shen2014}. If disc accretion is significantly delayed with respect to disc formation, we thus expect a solution a la \citet{cannizzo1990}, that predicts an accretion rate declining as $t^{-4/3}$. If on the other hand accretion already occurs while the disc is forming, we expect at late times a $t^{-5/3}$ decline in the accretion rate. Owing to a progressive disc formation, we consider the latter scenario to be a possibility, which seems favoured if cooling is inefficient, since $\tvisc < \tcirc$. 

The simulations presented in this paper allowed to get insight into the circularization process during TDEs. Several extensions of this work are possible including simulations considering a more eccentric star, the treatment of the interaction between the debris and the radiation emitted during the disc formation and the effect of the black hole spin.

\section*{Acknowledgments}

We acknowledge the use of SPLASH \citep{price2007} for generating the figures. C.B. is grateful to Inti Pelupessi for his help in implementing the particle splitting technique.

\label{lastpage}

\bibliography{biblio}

\end{document}